\documentclass[12pt,a4paper]{article}
\usepackage[utf8]{inputenc}
\usepackage{epsf}
\usepackage{latexsym,amssymb,euscript}
\usepackage[dvips]{graphicx}
\usepackage[numbers,sort&compress]{natbib}
\usepackage{amsmath}
\usepackage{slashed}
\usepackage{booktabs}
\usepackage{hyperref}
\usepackage{braket}
\usepackage{chngcntr}
\usepackage{bbold}
\usepackage{graphics}
\usepackage{graphicx}
\usepackage{caption}
\usepackage{subcaption}
\usepackage{pdfpages}
\usepackage{soul}
\usepackage[titletoc]{appendix}
\graphicspath{{./figures/}}
\hypersetup{
 linktocpage = true,
 urlcolor = purple,
 colorlinks = true,
 linkcolor = purple,
 anchorcolor = purple,
 citecolor = purple,
 pdfstartview = {XYZ null null 1.25} 
           }
\usepackage[left=2cm, right=2cm]{geometry}
\usepackage{pstricks}
\usepackage{color}
\usepackage{float}
\usepackage{academicons}
\definecolor{orcidlogocol}{HTML}{A6CE39}
\usepackage{fancyhdr}
\newcommand{\drv}{{\rm d}}
\newcommand{\prodscal}{\text{\large{\textbf{\textperiodcentered}}}}

\newcommand{\CnNLA}{{\cal C}_n^{\rm NLA}}
\newcommand{\CnDGLAP}{{\cal C}_n^{\rm DGLAP}}
\newcommand{\MSb}{\overline{\rm MS}}

\newcommand{\tcite}[1]{~\cite{#1}}
\newcommand{\tref}[1]{~\ref{#1}}

\newcommand{\orcidFGC}{\href{https://orcid.org/0000-0003-3299-2203}{\includegraphics[scale=0.1]{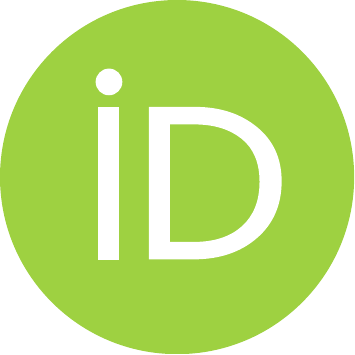}}}

\newcommand{\orcidMF}{\href{https://orcid.org/0000-0002-2408-2210}{\includegraphics[scale=0.1]{figures/logo-orcid.pdf}}}

\newcommand{\orcidDYI}{\href{https://orcid.org/0000-0001-5701-4364}{\includegraphics[scale=0.1]{figures/logo-orcid.pdf}}}

\newcommand{\orcidAP}{\href{https://orcid.org/0000-0001-8984-3036}{\includegraphics[scale=0.1]{figures/logo-orcid.pdf}}}

\begin{document}

\begin{titlepage}

\begin{center}
  {\LARGE \bf High-energy resummation in $\Lambda_c$ baryon production}
\end{center}

\vskip 0.5cm

\centerline{
Francesco~G.~Celiberto$^{\;1,2,3\;\dagger}$ \orcidFGC,
Michael~Fucilla$^{\;4,5\;\ddagger}$ \orcidMF,
Dmitry~Yu.~Ivanov$^{\;6\;\S}$ \orcidDYI,
and Alessandro~Papa$^{\;4,5\;\P}$ \orcidAP
}

\vskip .4cm

\centerline{${}^1$ {\sl European Centre for Theoretical Studies in Nuclear Physics and Related Areas (ECT*),}}
\centerline{\sl I-38123 Villazzano, Trento, Italy}
\vskip .18cm
\centerline{${}^2$ {\sl Fondazione Bruno Kessler (FBK), 
I-38123 Povo, Trento, Italy} }
\vskip .18cm
\centerline{${}^3$ {\sl INFN-TIFPA Trento Institute of Fundamental Physics and Applications,}}
\centerline{\sl I-38123 Povo, Trento, Italy}
\vskip .18cm
\centerline{${}^4$ {\sl Dipartimento di Fisica, Universit\`a della Calabria,}}
\centerline{\sl I-87036 Arcavacata di Rende, Cosenza, Italy}
\vskip .18cm
\centerline{${}^5$ {\sl Istituto Nazionale di Fisica Nucleare, Gruppo collegato di Cosenza,}}
\centerline{\sl I-87036 Arcavacata di Rende, Cosenza, Italy}
\vskip .18cm
\centerline{${}^6$ {\sl Sobolev Institute of Mathematics, 630090 Novosibirsk,
    Russia}}
\vskip 0.50cm

\begin{abstract}
\vspace{0.50cm}
\hrule \vspace{0.50cm}
We present a study on inclusive emissions of a double $\Lambda_c$ or of a $\Lambda_c$ plus a light-flavored jet system as probe channels in the semi-hard regime of QCD. 
Our formalism relies on the so-called hybrid high-energy/collinear factorization, where the standard collinear description is supplemented by the $t$-channel resummation \emph{\`a la} BFKL of energy logarithms up to the next-to-leading accuracy.
We make use of the {\tt JETHAD} modular interface, suited to the analysis of different semi-hard reactions, employing the novel {\tt KKSS19} parameterization for the description of parton fragmentation into $\Lambda_c$ baryons.
We provide predictions for rapidity distributions and azimuthal correlations, that can be studied at current and forthcoming LHC configurations, hunting for possible stabilizing effects of the high-energy series.
\vspace{0.50cm} \hrule
\vspace{0.50cm}
{
 \setlength{\parindent}{0pt}
}
\end{abstract}

\vfill
$^{\dagger}${\it e-mail}:
\href{mailto:fceliberto@ectstar.eu}{fceliberto@ectstar.eu}

$^{\ddagger}${\it e-mail}:
\href{mailto:michael.fucilla@unical.it}{michael.fucilla@unical.it}

$^{\S}${\it e-mail}:
\href{mailto:d-ivanov@math.nsc.ru}{d-ivanov@math.nsc.ru}

$^{\P}${\it e-mail}:
\href{alessandro.papa@fis.unical.it}{alessandro.papa@fis.unical.it}

\end{titlepage}

\section{Hors d'{\oe}uvre}
\label{introd}

Fixed-order perturbative QCD is a fundamental and successful approach for the description of strong-interaction processes ruled by one or more {\em hard} kinematic scales, {\it i.e.} scales much larger than 
$\Lambda_{\rm QCD}$. In some peculiar kinematic regimes, however, 
fixed-order perturbative calculations are no longer adequate to build sensible predictions and all-order resummations of some definite classes of contributions to the perturbative series become mandatory. One such kinematic regime is the \emph{semi-hard} one~\cite{Gribov:1984tu}, characterized by a center-of-mass energy, $\sqrt{s}$, much larger than the hard scale(s) of the process and, therefore, of the QCD mass scale. In this regime, large energy logarithms enter the perturbative series with a power which increases with the perturbative order, thus systematically compensating the smallness of the coupling and calling for all-order resummation. The approximation where only leading-order energy logarithms are resummed is called LLA; if also next-to-leading logarithms are resummed, we have the next-to-leading approximation or NLA. The LHC, with its world-record beam energies, is the best possible stage for semi-hard processes and a unique opportunity for testing theoretical approaches based on energy resummations in pertubative QCD.

A procedure for the systematic inclusion of large energy logarithms
to all orders in perturbative QCD, both in the LLA and in the NLA, is offered by the Balitsky--Fadin--Kuraev--Lipatov
(BFKL)~\cite{Fadin:1975cb,Kuraev:1976ge,Kuraev:1977fs,Balitsky:1978ic}. In this framework, the cross sections of hadronic processes take a peculiar factorized form, given by the convolution of two impact factors, related to the fragmentation of each
colliding particle to an identified final-state object, and a
process-independent Green's function. The evolution of the BFKL Green's function is built out of an integral equation, whose kernel is known at the next-to-leading order (NLO) both for forward scattering ({\it i.e.} for $t=0$ and color singlet in the $t$-channel)~\cite{Fadin:1998py,Ciafaloni:1998gs} and for any fixed, not growing with $s$, momentum transfer $t$ and any possible two-gluon color state in the $t$-channel~\cite{Fadin:1998jv,
Fadin:2000kx,Fadin:2000hu,Fadin:2004zq,Fadin:2005zj}.

The number of reactions predictable in the BFKL approach is limited  by the paucity of available impact factors, only a few of them being
known with NLO accuracy: 1) colliding-parton (quarks and gluons) impact factors~\cite{Fadin:1999de,Fadin:1999df,Ciafaloni:1998hu,Ciafaloni:2000sq}, which represent the common basis for the calculation of the 2) forward-jet impact factor~\cite{Bartels:2001ge,Bartels:2002yj,Caporale:2011cc,Ivanov:2012ms,Colferai:2015zfa} and of the 3) forward light-charged hadron one~\cite{Ivanov:2012iv}, the impact factors for 4) $\gamma^*$ to light-vector-meson leading twist transition~\cite{Ivanov:2004pp}, 5) $\gamma^*$ to $\gamma^*$ transition~\cite{Bartels:2000gt,Bartels:2001mv,Bartels:2002uz,Bartels:2003zi,Bartels:2004bi,Fadin:2001ap,Fadin:2002tu,Balitsky:2012bs}, and 6) proton to Higgs transition~\cite{Hentschinski:2020tbi}.

Combining pairwise the available NLO impact factors, a number of 
semi-hard (inclusive) reactions can be described within the BFKL approach in the NLA and predictions can be formulated, mainly in the form of azimuthal correlations or transverse momentum distributions, most of them accessible to current experiments at the LHC. These  reactions include (see also Ref.~\cite{Celiberto:2017ius}) the diffractive leptoproduction of two light vector mesons~\cite{Ivanov:2004pp,Ivanov:2005gn,Ivanov:2006gt,Enberg:2005eq}, the inclusive hadroproduction of two jets featuring large transverse momenta and well separated in rapidity (Mueller--Navelet channel~\cite{Mueller:1986ey}),
for which several phenomenological analyses have appeared so far~\cite{Colferai:2010wu,Caporale:2012ih,Ducloue:2013hia,Ducloue:2013bva,Caporale:2013uva,Ducloue:2014koa,Caporale:2014gpa,Ducloue:2015jba,Caporale:2015uva,Celiberto:2015yba,Celiberto:2015mpa,Celiberto:2016ygs,Celiberto:2016vva,Caporale:2018qnm}, 
the inclusive detection of two light-charged rapidity-separated
hadrons~\cite{Celiberto:2016hae,Celiberto:2016zgb,Celiberto:2017ptm,Celiberto:2017ydk}
or of a rapidity-separated pair formed by a light-charged hadron and a jet~\cite{Bolognino:2018oth,Bolognino:2019yqj,Bolognino:2019cac},
the inclusive production of rapidity-separated $\Lambda$-$\Lambda$ or $\Lambda$-jet pairs~\cite{Celiberto:2020rxb}.
For all these hadroproduction channels a \emph{hybrid} high-energy/collinear factorization was built up, where collinear parton distribution functions (PDFs) enter the definition of the BFKL impact factors.

If one waives the full NLA BFKL treatment, in favor of a partial inclusion of NLA resummation effects, then also LO impact factors come into play and a plethora of new semi-hard channels open up, such as three- and four-jet hadroproduction~\cite{Caporale:2015vya,Caporale:2015int,Caporale:2016soq,Caporale:2016vxt,Caporale:2016xku,Celiberto:2016vhn,Caporale:2016djm,Caporale:2016lnh,Caporale:2016zkc}, $J/\Psi$-jet~\cite{Boussarie:2017oae}, forward Drell--Yan dilepton production in association with a backward-jet tag~\cite{Golec-Biernat:2018kem}, 
heavy-quark pair photoproduction~\cite{Celiberto:2017nyx,Bolognino:2019ouc} and hadroproduction~\cite{Bolognino:2019yls,Bolognino:2021mrc}, Higgs-jet production~\cite{Celiberto:2020tmb}.

Moreover, the convolution of one impact factor with the \emph{unintegrated gluon distribution} (UGD) in the proton gives access to single-forward processes, such as the exclusive light vector-meson electroproduction~\cite{Anikin:2009bf,Anikin:2011sa,Besse:2013muy,Bolognino:2018rhb,Bolognino:2018mlw,Bolognino:2019bko,Bolognino:2019pba,Celiberto:2019slj}, the exclusive quarkonium photoproduction\tcite{Bautista:2016xnp,Garcia:2019tne,Hentschinski:2020yfm} and the inclusive tag of Drell--Yan pairs in forward directions\tcite{Motyka:2014lya,Brzeminski:2016lwh,Motyka:2016lta,Celiberto:2018muu}.
Starting from high-energy ingredients encoded in the UGD, first determinations of small-$x$ enhanced collinear PDFs\tcite{Ball:2017otu,Abdolmaleki:2018jln,Bonvini:2019wxf} as well as $T$-even gluon transverse-momentum-dependent distributions (TMDs)\tcite{Bacchetta:2020vty,Celiberto:2021zww} were recently proposed.

In this paper we propose the inclusive production of $\Lambda_c$ baryons in semi-hard regimes as a further probe of the high-energy resummation in kinematic ranges typical of current and upcoming experimental studies at the LHC. In particular, we focus on final states featuring the emission of a forward $\Lambda_c$ particle accompanied by a backward $\Lambda_c$ (panel a) of Fig.~\ref{fig:process}) or by a jet (panel b) of Fig.~\ref{fig:process}). The final-state inclusiveness is warranted by the emission of undetected hard partons strongly ordered in rapidity, while energy scales at play are large enough to make a variable-flavor number-scheme (VFNS) description valid (see, \emph{e.g.}, Refs.\tcite{Buza:1996wv,Bierenbaum:2009mv}). $\Lambda_c$ production at the LHC in central-rapidity ALICE and in forward-rapidity LHCb regimes was considered in Ref.\tcite{Maciula:2018iuh} within $k_T$-factorization.

Similar final-state configurations have been considered with NLA accuracy in~\cite{Celiberto:2020rxb}, with $\Lambda$ hyperons playing the role of the $\Lambda_c$. The motivations brought 
in Ref.~\cite{Celiberto:2020rxb} in support of the $\Lambda$ hyperon case plainly extend to the $\Lambda_c$. What makes the latter case interesting on its own is the following consideration. It is well known that NLA BFKL suffers from severe instabilities, due to 
being the NLO corrections in the BFKL kernel and in impact factors large in size and opposite in sign with respect to the LO. In 
Ref.~\cite{Celiberto:2020rxb} and in our previous NLA BFKL studies the procedure adopted to tame this instability was the use of the 
MOM scheme for the strong coupling renormalization with Brodsky--Lepage--Mackenzie (BLM) optimization~\cite{Brodsky:1996sg,Brodsky:1997sd,Brodsky:1998kn,Brodsky:2002ka} 
of the renormalization scale fixing. This allowed to obtain stable results for cross sections and azimuthal correlations or ratios between them, at the price of fixing the renormalization scale at values much larger than the kinematic ones, typical of the  considered process. The interesting feature of the processes suggested in this paper is that the BLM optimization scales turn out to be much smaller than in previous studies, due to a subtle interplay between proton PDFs and $\Lambda_c$ fragmentation functions (FFs), discussed in details below. The obvious interpretation of this phenomenon is that $\Lambda_c$ production, due to the particular shape of the corresponding FFs, leads to an improved stability of the BFKL series,
making the suggested processes worthy of attention in LHC experimental analyses. 

The paper is organized as follows: in Section~\ref{theory} we introduce our theoretical setup, in Section\tref{pheno} we show and discuss results of our numerical analysis, in Section\tref{conclusions} we draw our conclusions and propose future prospects.

\section{Hybrid factorization for $\Lambda_c$ production}
\label{theory}

\subsection{NLA-resummed cross section}
\label{NLA}

\begin{figure}[t]
\centering
\includegraphics[width=0.45\textwidth]{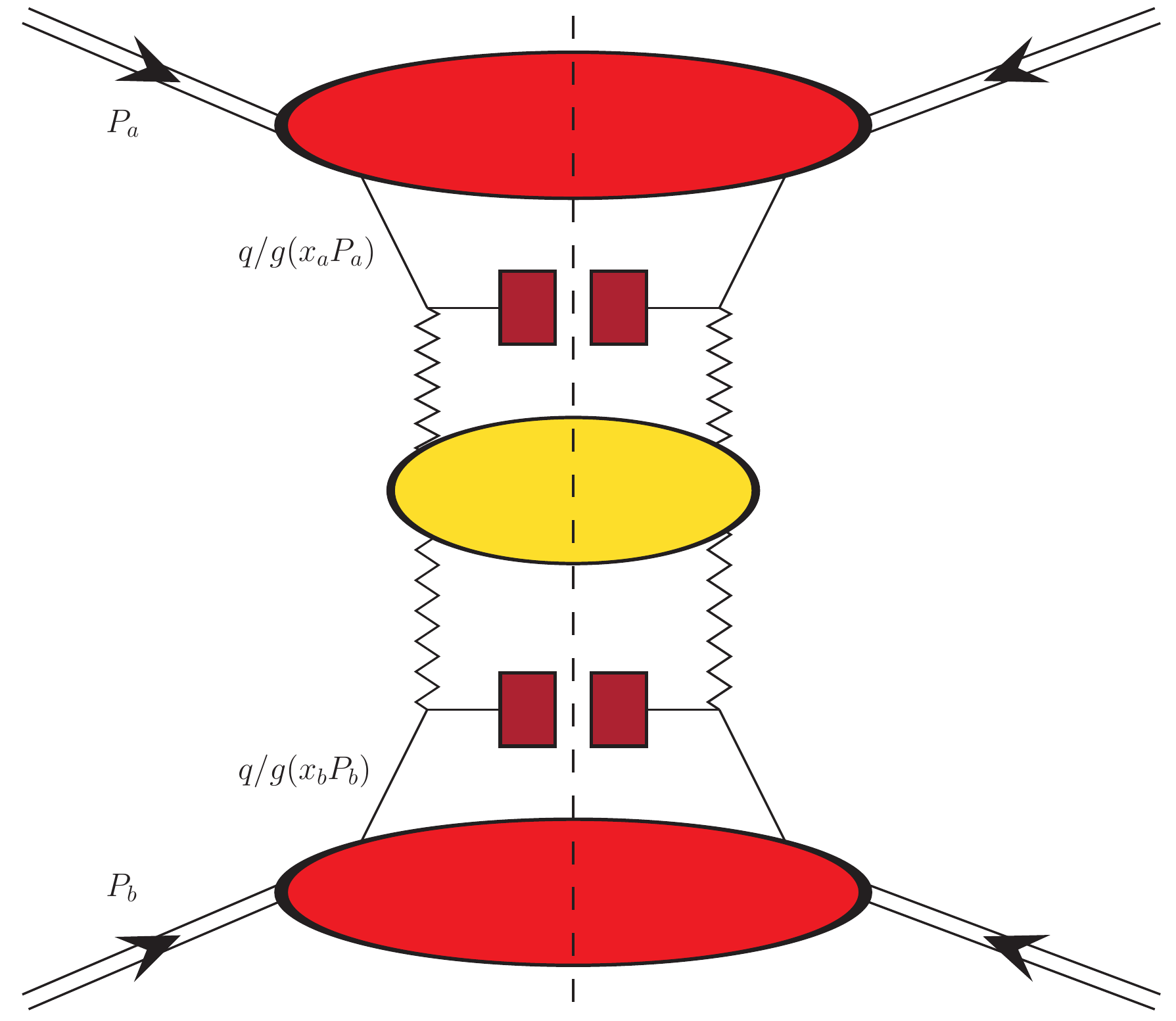}
\hspace{1.40cm}
\includegraphics[width=0.45\textwidth]{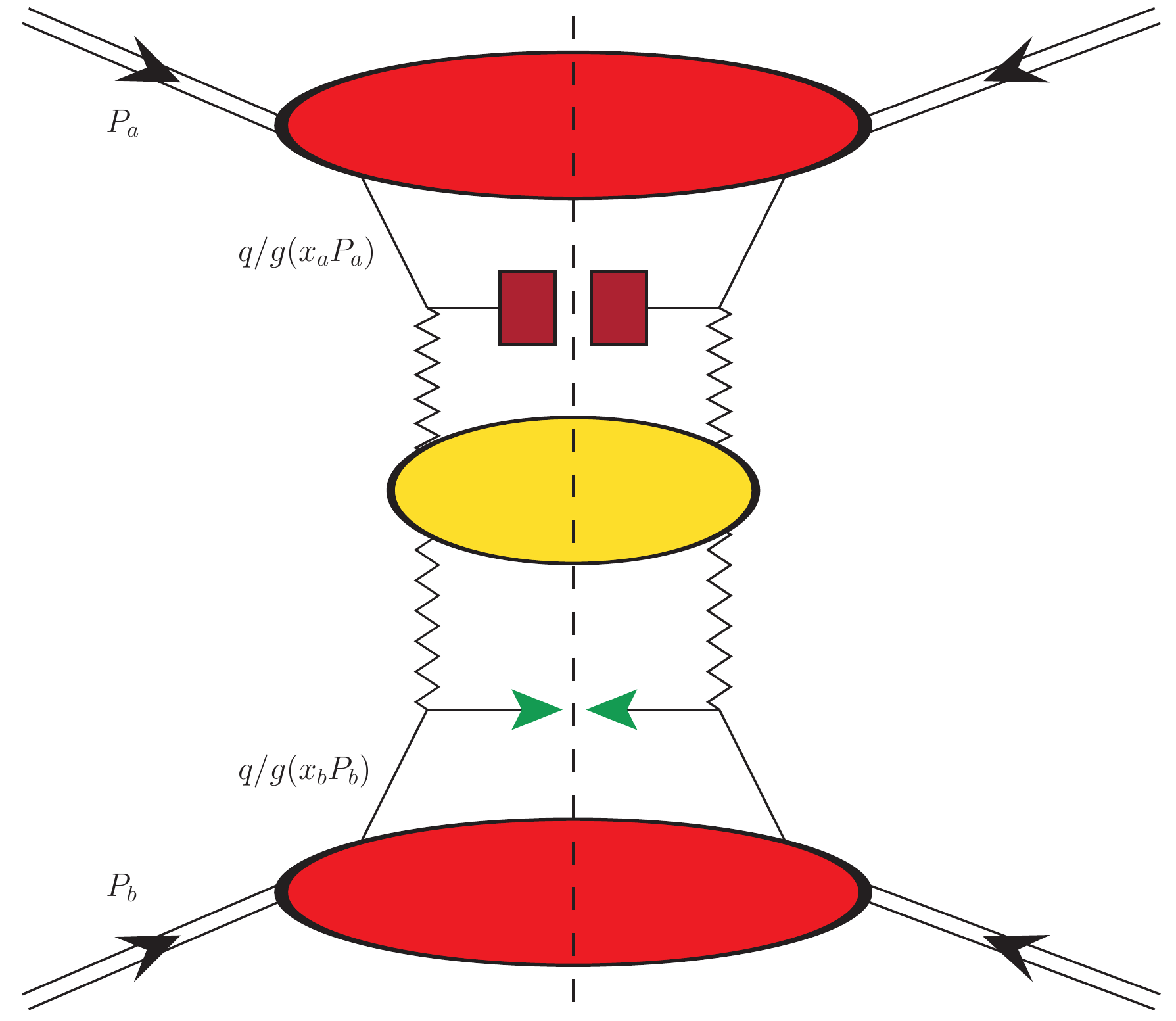}
\\ \vspace{0.25cm}
\hspace{-0.50cm}
a) Double $\Lambda_c$ \hspace{7.00cm}
b) $\Lambda_c$ $+$ jet
\caption{Diagrammatic representation of a) the double $\Lambda_c$ and of b) the $\Lambda_c$ $+$ jet production in hybrid high-energy/collinear factorization. Red blobs stand for proton collinear PDFs, whereas bordeaux rectangles denote baryon collinear FFs and green arrows refer to the jet selection function. The BFKL gluon Green's function, represented by the yellow central blob, is connected to impact factors through Reggeon (zigzag) lines. 
Diagrams were realized by making use of the {\tt JaxoDraw 2.0} interface~\cite{Binosi:2008ig}.}
\label{fig:process}
\end{figure}

The two hadronic reactions object of our investigation are (see Fig.~\ref{fig:process}):
\begin{equation}
\label{process_LL}
 {\rm proton}(P_a) + {\rm proton}(P_b) \to \Lambda_c^\pm(p_1, y_1) + X + \Lambda_c^\pm(p_2 , y_2) \;,
\end{equation}
\begin{equation}
\label{process_LJ}
 {\rm proton}(P_a) + {\rm proton}(P_b) \to \Lambda_c^\pm(p_1, y_1) + X + {\rm jet}(p_2 , y_2) \;,
\end{equation}
where a $\Lambda_c$ particle (we are inclusive on the baryon charge) is accompanied by another $\Lambda_c$ or by a light-flavored jet\footnote{Since we are working in the VFNS, where all parton quarks are treated as light ones, the jet can be generated by any quark, up to the active flavor number, $n_f$, or by a gluon.}. Produced objects have transverse momenta, $|\vec p_{1,2}| \gg \Lambda_{\rm QCD}$, large enough to justify the use of pQCD. At the same time, their large distance in rapidity, $\Delta Y \equiv y_1 - y_2$, ensures diffractive final states. An undetected gluon activity, denoted as $X$, accompanies many-particle emissions. 
We will consider values of $\Delta Y$ larger than zero, so that the 1-labeled $\Lambda_c$ will be always a forward particle, while the 2-labeled object will be always a backward one.
We take as Sudakov basis the one generated by momenta of incoming protons, $P_{a,b}$, thus having $P_{a,b}^{2} = 0$ and $({P_a} \prodscal {P_b}) = s/2$, with $s$ the hadronic center-of-mass energy squared.
In this way, one can decompose the final-state transverse momenta on that basis:
\begin{equation}
\label{Sudakov}
 p_{1,2} = x_{1,2} P_{a,b} + \frac{\vec p_{1,2}^{\,2}}{x_{1,2} s} P_{b,a} + p_{{1,2 \perp}} \;, \qquad p_{1,2 \perp}^2 \equiv - \vec p_{1,2}^{\,2} \;,
\end{equation}
where the subscript $O_1$ refers to the first emitted object (always a baryon) and $O_2$ refers to the second one (another baryon or a jet).
In the center-of-mass frame, the following relations between longitudinal fractions, rapidities, and transverse momenta of emitted particles hold
\begin{equation}
\label{xyp}
 x_{1,2} = \frac{|\vec p_{1,2}|}{\sqrt{s}} e^{\pm y_{1,2}} \;, \qquad \drv y_{1,2} = \pm \frac{d x_{1,2}}{x_{1,2}} \;,
 \qquad \Delta Y = y_1 - y_2 = \ln\frac{x_1 x_2 s}{|\vec p_1| |\vec p_2|} \;.
\end{equation}

A pure fixed-order treatment relies on the standard collinear factorization, where the on-shell partonic cross section is convoluted with PDFs and FFs.
In the double $\Lambda_c$ channel (panel a) of Fig.~\ref{fig:process}) one has at LO
\begin{align}
\label{sigma_collinear_LL} \nonumber
 \frac{\drv \sigma^{[pp \,\to\, \Lambda_c \Lambda_c]}_{\rm coll.}}{\drv x_1 \drv x_2 \drv^2 \vec p_1 \drv^2 \vec p_2}
 =\sum_{a,b} &\int_0^1 \drv x_a \int_0^1 \drv x_b\ 
 f_a\left(x_a\right) f_b\left(x_b\right)
\\ 
 \times \, &\int_{x_1}^1 \frac{\drv z_1}{z_1} \int_{x_2}^1 \frac{\drv z_2}{z_2}\
 D^\Lambda_a\left(\frac{x_1}{z_1}\right) D^\Lambda_b\left(\frac{x_2}{z_2}\right)
 \frac{\drv {\hat\sigma}_{a,b}\left(\hat s\right)}
 {\drv x_a \drv x_b \drv z_1 \drv z_2 \drv^2 \vec p_1 \drv^2 \vec p_2}\;,
\end{align}
where the ($a,b$) indices run over all quark and antiquark flavors and the gluon, $f_{a,b}\left(x\right)$ are the incoming protons' PDFs, $D^\Lambda_{a,b}\left(\frac{x}{z}\right)$ are the outgoing $\Lambda_c$ FFs, $x_{a,b}$ are the longitudinal fractions of the struck partons, $z_{1,2}$ are the longitudinal fractions of partons fragmenting into baryons, and $\drv \hat\sigma_{a,b}\left(\hat s \right)$ is the partonic cross section, with $\hat s \equiv x_ax_bs$ the partonic center-of-mass energy squared.
Analogously, in the $\Lambda_c$ plus jet channel (panel b) of Fig.~\ref{fig:process}) one has at LO
\begin{align}
\label{sigma_collinear_LJ} \nonumber
 \frac{\drv \sigma^{[pp \,\to\, \Lambda_c \, {\rm jet}]}_{\rm coll.}}{\drv x_1 \drv x_2 \drv^2 \vec p_1 \drv^2 \vec p_2}
 =\sum_{a,b} &\int_0^1 \drv x_a \int_0^1 \drv x_b\ 
 f_a\left(x_a\right) f_b\left(x_b\right)
\\ 
 \times \, &\int_{x_1}^1 \drv z_1\
 D^\Lambda_a\left(\frac{x_1}{z_1}\right)
 \frac{\drv {\hat\sigma}_{a,b}\left(\hat s\right)}
 {\drv x_a \drv x_b \drv z_1 \drv x_2 \drv^2 \vec p_1 \drv^2 \vec p_2}\;.
\end{align}
For the sake of readability, the explicit dependence of PDFs, FFs, and the partonic cross section on the factorization scale has been dropped everywhere.

At variance with the standard collinear approach, we build our hybrid setup by starting from the high-energy factorization which naturally emerges inside the BFKL formalism, and then we improve our description by plugging collinear ingredients in.
It is convenient to write the cross section as a Fourier series of the azimuthal-angle coefficients, ${\cal C}_{n \ge 0}$, this leading to the following general expression
\begin{equation}
 \label{dsigma_Fourier}
 \frac{\drv \sigma}{\drv y_1 \drv y_2 \drv \vec p_1 \drv \vec p_2 \drv \varphi_1 \drv \varphi_2} =
 \frac{1}{(2\pi)^2} \left[{\cal C}_0 + 2 \sum_{n=1}^\infty \cos (n \varphi)\,
 {\cal C}_n \right]\, ,
\end{equation}
where $\varphi_{1,2}$ are the azimuthal angles of the tagged objects and $\varphi \equiv \varphi_1 - \varphi_2 - \pi$.
The BFKL approach provides us with a consistent definition of NLA-resummed azimuthal coefficients, ${\cal C}_n = \CnNLA$, whose definition in the $\MSb$ renormalization scheme reads (details on the derivation can be found in Ref.~\cite{Caporale:2012ih})
\[
 \CnNLA \equiv \int_0^{2\pi} \drv \varphi_1\int_0^{2\pi} \drv \varphi_2\,
 \cos (n \varphi) \,
 \frac{\drv \sigma_{\rm NLA}}{\drv y_1 \drv y_2\, \drv |\vec p_1| \, \drv |\vec p_2| \drv \varphi_1 \drv \varphi_2}\;
\]
\[
 = \frac{e^{\Delta Y}}{s} 
 \int_{-\infty}^{+\infty} \drv \nu \, e^{{\Delta Y} \bar \alpha_s(\mu_R)\left\{\chi(n,\nu)+\bar\alpha_s(\mu_R)
 \left[\bar\chi(n,\nu)+\frac{\beta_0}{8 N_c}\chi(n,\nu)\left[-\chi(n,\nu)+\frac{10}{3}+4\ln\left(\frac{\mu_R}{\sqrt{|\vec p_1| |\vec p_2|}}\right)\right]\right]\right\}}
\]
\[
 \times \, \alpha_s^2(\mu_R) c_1(n,\nu,|\vec p_1|, x_1)[c_2(n,\nu,|\vec p_2|,x_2)]^*\,
\]
\begin{equation}
\label{Cn_MSb}
 \times \, \left\{1
 +\alpha_s(\mu_R)\left[\frac{\hat c_1(n,\nu,|\vec p_1|,x_1)}{c_1(n,\nu,|\vec p_1|,x_1)}
 +\left[\frac{\hat c_2(n,\nu,|\vec p_2|, x_2)}{c_2(n,\nu,|\vec p_2|,x_2)}\right]^*
 +  \bar \alpha_s(\mu_R) 
 \, \Delta Y
 \frac{\beta_0}{4 \pi}\chi(n,\nu)f(\nu)
 \right]
 \right\} \;,
\end{equation}
where $\bar \alpha_s(\mu_R) \equiv \alpha_s(\mu_R) N_c/\pi$, with $N_c$ the color number and $\beta_0 = 11N_c/3 - 2 n_f/3$ the first coefficient of the QCD $\beta$-function.
The expression in Eq.~(\ref{Cn_MSb}) for LO BFKL eigenvalues reads
\begin{equation}
 \label{kernel_LO}
 \chi\left(n,\nu\right) = -2\gamma_{\rm E} - \psi\left(\frac{n}{2} + \frac{1}{2} + i \nu \right) - \psi\left(\frac{n}{2} + \frac{1}{2} - i\nu \right) \, ,
\end{equation}
with $\gamma_{\rm E}$ the Euler--Mascheroni constant and $\psi(z) \equiv \Gamma^\prime
(z)/\Gamma(z)$ the logarithmic derivative of the Gamma function, whereas $\bar\chi(n,\nu)$ comes from the NLO correction to the BFKL kernel and was calculated in Ref.~\cite{Kotikov:2000pm} (see also Ref.~\cite{Kotikov:2002ab}).
Then, the $c_{1,2}$ functions stand for the LO impact factors, calculated in the Mellin space, describing the emissions of forward/backward objects. The LO impact factor for the production of a $\Lambda_c$ baryon reads
\[
c_\Lambda(n,\nu,|\vec p\,|,x) 
= 2 \sqrt{\frac{C_F}{C_A}}
(|\vec p\,|^2)^{i\nu-1/2}\,\int_{x}^1\frac{\drv z}{z}
\left( \frac{z}{x} \right)
^{2 i\nu-1} 
\]
\begin{equation}
\label{LOLIF}
 \times \left[\frac{C_A}{C_F}f_g(z)D_g^\Lambda\left(\frac{x}{z}\right)
 +\sum_{a=q,\bar q}f_a(z)D_a^\Lambda\left(\frac{x}{z}\right)\right] \;.
\end{equation}
Similarly, the light jet is portrayed by the corresponding LO impact factor
\begin{equation}
 \label{LOJIF}
 c_J(n,\nu,|\vec p\,|,x) =  2 \sqrt{\frac{C_F}{C_A}}
 (|\vec p\,|^2)^{i\nu-1/2}\,\left(\frac{C_A}{C_F}f_g(x)
 +\sum_{b=q,\bar q}f_b(x)\right) \;.
\end{equation}
Furthermore, the $f(\nu)$ function is defined as
\begin{equation}
 f(\nu) = \frac{1}{2} \left[ i \frac{\drv}{\drv \nu} \ln\left(\frac{c_1}{c_2^*}\right) + 2 \ln\left(|\vec p_1| |\vec p_2|\right) \right] \;.
\label{fnu}
\end{equation}
The remaining quantities in Eq.~(\ref{Cn_MSb}), $\hat c_{1,2}$, are the NLO corrections to the hadron impact factor, whose analytic expression was calculated in Ref.\tcite{Ivanov:2012iv}, and the jet impact-factor one. For this one we will employ a simple version (see Ref.\tcite{Ivanov:2012ms}), suited to numerical studies, which encodes a jet selection function calculated in the ``small-cone'' approximation (SCA)~\cite{Furman:1981kf,Aversa:1988vb}.

Eqs.~(\ref{Cn_MSb}),~(\ref{LOLIF}),~and~(\ref{LOJIF}) clearly show how our hybrid factorization is realized. The cross section is high-energy factorized in terms of gluon Green's function and impact factors, and the latter ones embody collinear PDFs and FFs. We remark that the description of $\Lambda_c$ particles in terms of light-hadron impact factors~(Eq.~(\ref{LOLIF})) is adequate, provided that energy scales are much larger than the $\Lambda_c$ mass. This condition is guaranteed by the transverse-momentum ranges of our interest (see Section\tref{observables}).

By truncating the NLA azimuthal coefficients in Eq.~(\ref{Cn_MSb}) to the ${\cal O}(\alpha_s^3)$, we obtain a fixed-order formula that acts an effective high-energy DGLAP counterpart of our BKFL-resummed expression. Thus, we keep the leading-power asymptotic pattern of a pure NLO DGLAP description, eliminating at the same time those terms which are suppressed by inverse powers of $\hat s$.
Our DGLAP formula can be cast in the form
\begin{equation}
\label{Cn:dglap}
 \CnDGLAP = \frac{e^{\Delta Y}}{s}
 \int_{-\infty}^{+\infty} \drv \nu \, \alpha_s(\mu_R)^2
 c_1(n,\nu,|\vec p_1|,x_1)[c_2(n,\nu,|\vec p_2|,x_2)]^*\,
\end{equation}
\[
 \times \, \left\{1+\alpha_s(\mu_R)\left[\Delta Y \frac{C_A}{\pi} \chi(n,\nu)+\frac{\hat c_1(n,\nu,|\vec p_1|,x_1)}{c_1(n,\nu,|\vec p_1|, x_1)}
 +\left[\frac{\hat c_2(n,\nu,|\vec p_2|, x_2)}{c_2(n,\nu,|\vec p_2|,x_2)}\right]^* 
 \right]
 \right\} \;,
\]
where an expansion up to terms proportional to $\alpha_s(\mu_R)$ replaces the BFKL exponentiated kernel.

Starting from Eqs.~(\ref{Cn_MSb})~and~(\ref{Cn:dglap}), it is possible to obtain corresponding expressions in the MOM scheme by performing the finite renormalization
\begin{equation}
 \label{as_MOM}
 \alpha_s^{\MSb}(\mu_R) \;\to\;
 \alpha^{\rm MOM}_s (\mu_R) = 
 - \frac{\pi}{2 (T^{\beta} + T^{\rm conf})}
 \left( 1 - \sqrt{1 + 4 \, \alpha^{(\overline{\rm MS})}_s (\mu_R) \frac{T^{\beta} + T^{\rm conf}}{\pi}} \right) \;,
\end{equation}
where
\begin{equation}
\label{T_bc}
T^{\beta}=-\frac{\beta_0}{2}\left( 1+\frac{2}{3}I \right) \;, \quad 
T^{\rm conf}= \frac{C_A}{8}\left[ \frac{17}{2}I +\frac{3}{2}\left(I-1\right)\xi
+\left( 1-\frac{1}{3}I\right)\xi^2-\frac{1}{6}\xi^3 \right] \; ,
\end{equation}
with $I = -2 \int_0^1 \drv \delta \frac{\ln \delta}{1 - \delta + \delta^2} \simeq 2.3439$ and $\xi$ a gauge parameter, fixed at zero in the following.

\subsection{BLM scale optimization}
\label{BLM}

According to the BLM method, the \emph{optimal} renormalization-scale value, labeled as $\mu_R^{\rm BLM}$, is the value of $\mu_R$ that cancels the non-conformal, $\beta_0$-dependent part of a given observable.
A suitable procedure, recently set up\tcite{Caporale:2015uva}, allows us to remove all non-conformal contributions that appear both in the NLA BFKL kernel and in the NLO non-universal impact factors of high-energy distributions. Its application makes $\mu_R^{\rm BLM}$ dependent on the energy of the process and thus on $\Delta Y$. 

As a preliminary step, we need to perform a finite renormalization from the $\overline{\rm MS}$ scheme to
the MOM one (see Eq.~(\ref{as_MOM})).
Thus, the condition for the BLM scale setting for a given azimuthal coefficient, $C_n$, is given as a solution of the integral equation
\begin{equation}
\label{Cn_beta_int}
  C_n^{[\beta]}(s, \Delta Y) = 
  \int \drv \Phi(y_{1,2}, |\vec p_{1,2}|, \Delta Y) \,
  \, {\cal C}_n^{[\beta]}  = 0 \, ,
\end{equation}
with $\drv \Phi(y_{1,2}, |\vec p_{1,2}|, \Delta Y)$ the final-state differential phase space (see Section\tref{observables}),
\[
 {\cal C}^{[\beta]}_n
 \propto \!\!
 \int^{\infty}_{-\infty} \!\!\drv\nu\,e^{\Delta Y \bar \alpha^{\rm MOM}_s(\mu^{\rm BLM}_R)\chi(n,\nu)}
 c_1(n,\nu,|\vec p_1|,x_1)[c_2(n,\nu,|\vec p_2|,x_2)]^*
\]
\begin{equation}
\label{Cn_beta}
 \times \, \left[{\upsilon}(\nu) + \bar \alpha^{\rm MOM}_s(\mu^{\rm BLM}_R) \Delta Y \: \frac{\chi(n,\nu)}{2} \left(- \frac{\chi(n,\nu)}{2} + {\upsilon}(\nu) \right) \right] \, ,
\end{equation}
and
\begin{equation}
\label{upsilon_nu}
{\upsilon}(\nu) = f(\nu) + \frac{5}{3} + 2 \ln \left( \frac{\mu^{\rm BLM}_R}{\sqrt{|\vec p_1| |\vec p_2|}} \right) - 2 - \frac{4}{3} I \, .
\end{equation}
For the sake of convenience, we introduce the ratio between the BLM scale and the \emph{natural} one suggested by the kinematic of the process, namely
$\mu_N \equiv \sqrt{m_{1 \perp} m_{2 \perp}}$, so that $ C_{\mu}^{\rm BLM} \equiv \mu_R^{\rm BLM}/\mu_N$, and we look for values of $C_{\mu_R}$ that solve Eq.~(\ref{Cn_beta_int}). Here, $m_{i \perp}$ stands for the $i$-th particle transverse mass. Therefore, one always has  $m_{1 \perp} = \sqrt{|\vec p_1|^2 + m_{\Lambda_c}^2}$, with $ m_{\Lambda_c} = 2.286$ GeV. Then, $m_{2 \perp} = \sqrt{|\vec p_2|^2 + m_{\Lambda_c}^2}$ in the double $\Lambda_c$ production, while $m_{2 \perp}$ coincides with the jet $p_T$ in the other channel. We set $\mu_F = \mu_R$ everywhere, as assumed by most of the existent PDF parameterizations.

Finally, the BLM scale value is inserted into expressions for the integrated coefficients, and we get the following NLA BFKL formula in the MOM renormalization scheme
\[
C_n^{\rm NLA} = 
 \int \drv \Phi(y_{1,2}, |\vec p_{1,2}|, \Delta Y) \; 
 \frac{e^{\Delta Y}}{s} 
 \int_{-\infty}^{+\infty} \drv \nu \,
 \left(\alpha^{\rm MOM}_s (\mu^{\rm BLM}_R)\right)^2
\]
\[
 \times \,
 e^{\Delta Y \bar \alpha^{\rm MOM}_s(\mu^{\rm BLM}_R)\left[\chi(n,\nu)
 +\bar \alpha^{\rm MOM}_s(\mu^{\rm BLM}_R)\left(\bar \chi(n,\nu) +\frac{T^{\rm conf}}
 {3}\chi(n,\nu)\right)\right]}
 c_1(n,\nu,|\vec p_1|,x_1)[c_2(n,\nu,|\vec p_2|,x_2)]^*
\]
\begin{equation}
\label{Cn_NLA_int_blm}
\times \,
 \left\{1 + \alpha^{\rm MOM}_s(\mu^{\rm BLM}_R)\left[\frac{\bar c_1(n,\nu,|\vec p_1|,x_1)}{c_1(n,\nu,|\vec p_1|,x_1)}
 +\left[\frac{\bar c_2(n,\nu,|\vec p_2|, x_2)}{c_2(n,\nu,|\vec p_2|,x_2)}\right]^*
 +\frac{2T^{\rm conf}}{\pi} \right] \right\} \, ,
\end{equation}
where $\bar c_{{1,2}}(n,\nu,|\vec p_{1,2}|,x_{1,2})$ stand for the NLO impact-factor corrections after removing the $\beta_0$-dependent terms, which can be universally
expressed via the LO impact factors, $c_{1,2}$. One has
\begin{equation}
\label{IF_NLO_sub}
\bar c_{1,2} = \hat c_{1,2} + \frac{\beta_0}{4 N_c} \left[ \mp i \frac{\drv}{\drv \nu} c_{1,2} + \left( \ln \mu_R^2 + \frac{5}{3} \right) c_{1,2} \right] \, .
\end{equation}

Analogously, we expand and truncate to the ${\cal O}(\alpha_s^3)$ the BFKL kernel in Eq.~(\ref{Cn_NLA_int_blm}), this getting a BLM-MOM expression for the high-energy DGLAP limit
\[
C_n^{\rm DGLAP} = 
 \int \drv \Phi(y_{1,2}, |\vec p_{1,2}|, \Delta Y) \; 
 \frac{e^{\Delta Y}}{s} 
 \int_{-\infty}^{+\infty} \drv \nu \,
 \left(\alpha^{\rm MOM}_s (\mu^{\rm BLM}_R)\right)^2
 c_1(n,\nu,|\vec p_1|,x_1)[c_2(n,\nu,|\vec p_2|,x_2)]^*
\]
\begin{equation}
\label{Cn_DGLAP_int_blm}
\times \,
 \left\{1 + \alpha^{\rm MOM}_s(\mu^{\rm BLM}_R)\left[\Delta Y \frac{C_A}{\pi} \chi(n,\nu) + \frac{\bar c_1(n,\nu,|\vec p_1|,x_1)}{c_1(n,\nu,|\vec p_1|,x_1)}
 +\left[\frac{\bar c_2(n,\nu,|\vec p_2|, x_2)}{c_2(n,\nu,|\vec p_2|,x_2)}\right]^*
 +\frac{2T^{\rm conf}}{\pi} \right] \right\} \, .
\end{equation}
The corresponding formul{\ae} of Eqs.~(\ref{Cn_NLA_int_blm})~and~(\ref{Cn_DGLAP_int_blm}) in the $\MSb$ scheme are obtained by making the substitutions (note that the value of the renormalization scale is left unchanged)
\begin{equation}
 \label{MOM_to_MSb}
 \qquad \alpha_s^{\rm MOM}(\mu^{\rm BLM}_R) \,\to\, \alpha_s^{\MSb}(\mu^{\rm BLM}_R) \;, \quad T^{\rm conf} \,\to\, - T^\beta \;.
\end{equation}

\section{Numerical analysis}
\label{pheno}

\subsection{Rapidity and azimuthal distributions}
\label{observables}

Key ingredients for building our distributions are the azimuthal coefficients integrated over rapidity and transverse momenta of the two tagged object, their rapidity separation being kept fixed:
\begin{equation}
 \label{Cn_int}
 C_n =
 \int_{y_1^{\rm min}}^{y_1^{\rm max}} \drv y_1
 \int_{y_2^{\rm min}}^{y_2^{\rm max}} \drv y_2
 \int_{p_1^{\rm min}}^{p_1^{\rm max}} \drv |\vec p_1|
 \int_{p_2^{\rm min}}^{p_2^{\rm max}} \drv |\vec p_2|
 \, \,
 \delta (\Delta Y - y_1 + y_2)
 \, \,
 {\cal C}_n\left(|\vec p_1|, |\vec p_2|, y_1, y_2 \right)
 \, .
\end{equation}
The ${\cal C}_n$ and $C_n$ coefficients can refer to the corresponding NLA BFKL calculations (see Eq.~(\ref{Cn_MSb})) or the ones taken in the high-energy DGLAP limit (see Eq.~(\ref{Cn:dglap})).

In order to match realistic LHC configurations, we allow the rapidity of $\Lambda_c$ baryons to be in the range from $-2.0$ and $2.0$, while their transverse momentum goes from 10 GeV to $p_H^{\max} \simeq 21.5$ GeV. These cuts are borrowed from typical analyses of the $\Lambda_b$ particle at CMS\tcite{Chatrchyan:2012xg}, whereas the $p_H^{\max}$ value is constrained by the energy-scale lower cutoff on the FF sets (see Section\tref{numerics}). As for the jet, we consider standard CMS configurations\tcite{Khachatryan:2016udy}, namely $|y_J| < 4.7$ and 35 GeV $< p_J <$ 60 GeV.
A major benefit of allowing for jet detection also by the CASTOR ultra-backward detector ($-6.6 < y_J < -5.2$)\tcite{Khachatryan:2020mpd,Baur:2019yfg} is the possibility to test our observables on larger values of rapidity intervals, say $\Delta Y \lesssim 9$. 
However, in our previous work on $\Lambda$ hyperons\tcite{Celiberto:2020rxb} we have highlighted how, in this kinematics regime, large values of partons' longidudinal fractions effectively restrict the weight of the undetected gluon radiation. This leads to  the appearance of large Sudakov-type double logarithms (threshold double logarithms) in the perturbative series, that have to be resummed to all orders. Since this resummation mechanism has not been yet embodied in our approach, we postpone the investigation of $\Lambda_c$ emissions in CASTOR-jet configurations to future, dedicated studies.

The integrated coefficients defined in Eq.~(\ref{Cn_int}) permit us to study the $\varphi$-summed cross section, $C_0$, and the azimuthal-correlation ratios, $R_{nm} = C_m/C_m$, as functions of $\Delta Y$.
The $R_{n0}$ ratios have an immediate physical interpretation, being the moments $\langle \cos n \varphi \rangle$, while the ones without zero indices correspond to ratios of cosines, $\langle \cos n \varphi \rangle / \langle \cos m \varphi \rangle$~\cite{Vera:2006un,Vera:2007kn}. In our study we fix the center-of-mass energy at $\sqrt{s} = 13$~TeV.

\subsection{{\tt JETHAD} settings}
\label{numerics}

We performed our phenomenological studies by making use of the {\tt JETHAD} modular interface under development at our Group. Numerical computations of distributions as well as BLM scales were done via the \textsc{Fortran 2008} work package implemented in {\tt JETHAD}, whereas an automatized \textsc{Python 3.0} analyzer was developed for elaboration of results.

Collinear PDFs were calculated via the {\tt MMHT14} NLO PDF
set~\cite{Harland-Lang:2014zoa} as provided by the {\tt LHAPDFv6.2.1}
interpolator~\cite{Buckley:2014ana}, and a two-loop
running coupling with $\alpha_s\left(M_Z\right)=0.11707$ and a
dynamic-flavor threshold was chosen.
We described the parton fragmentation into $\Lambda_c$ baryons in terms of the novel {\tt KKSS19} NLO FF set\tcite{Kniehl:2020szu} (see also Refs.\tcite{Kniehl:2005de,Kniehl:2012ti}), whose native implementation was directly linked to {\tt JETHAD}. This parameterization mainly relies on a description \emph{\`a la} Bowler\tcite{Bowler:1981sb} for $c$ and $b$ quark/antiquark flavors. Technical details on the fitting procedure are presented in Section IV of Ref.\tcite{Kniehl:2020szu}. Here, we limit ourselves to saying that the use of a given VFNS PDF or FF set is valid in our approach, provided that energy scales at work are much larger than thresholds for DGLAP evolution of heavier quarks. Since our scales are always higher than 10 GeV (see Section\tref{observables}), while {\tt KKSS19} thresholds for $c$ and $b$ quarks are respectively 1.5 GeV and 5 GeV, this requirement is fulfilled.
Lighter-hadron emissions ($\Lambda$ hyperons, pions, kaons, and protons) in Figs.\tref{fig:BLM_scales_HSA} and\tref{fig:C0_HSA} were described in terms {\tt AKK08} NLO FFs\tcite{Albino:2008fy}, which are the closest in technology to {\tt KKSS19}.

Nested integrations over phase space, $\nu$ variable, and longitudinal fractions inside impact factors were mainly evaluated via an adaptative-quadrature strategy provided by the {\tt JETHAD} integrators. In all cases, their global uncertainty was kept below 1\%.

We performed a dedicated study on the sensitivity of our predictions on scale variation. More in particular, we gauged the effect of concurrently varying $\mu_R$ and $\mu_F$ around their \emph{natural} values and their BLM \emph{optimal} ones, in the range 1/2 to two.
The $C_{\mu}$ parameter entering the figures in Section\tref{discussion} stands for the ratio $C_\mu = \mu_{R,F}/\mu_N$. Error bands in all our plots show the combined uncertainty coming from numerical integration and scale variation, this latter being the dominant one.
All calculations were done in the $\MSb$ scheme.

An extension of our analysis that includes all systematic uncertainties together with a comparison of results obtained with different FF sets, as the {\tt DMS20} one\tcite{Delpasand:2020vlb}, is postponed to a future work.

\subsection{Discussion}
\label{discussion}

\begin{figure}[t]
\centering

   \includegraphics[scale=0.53,clip]{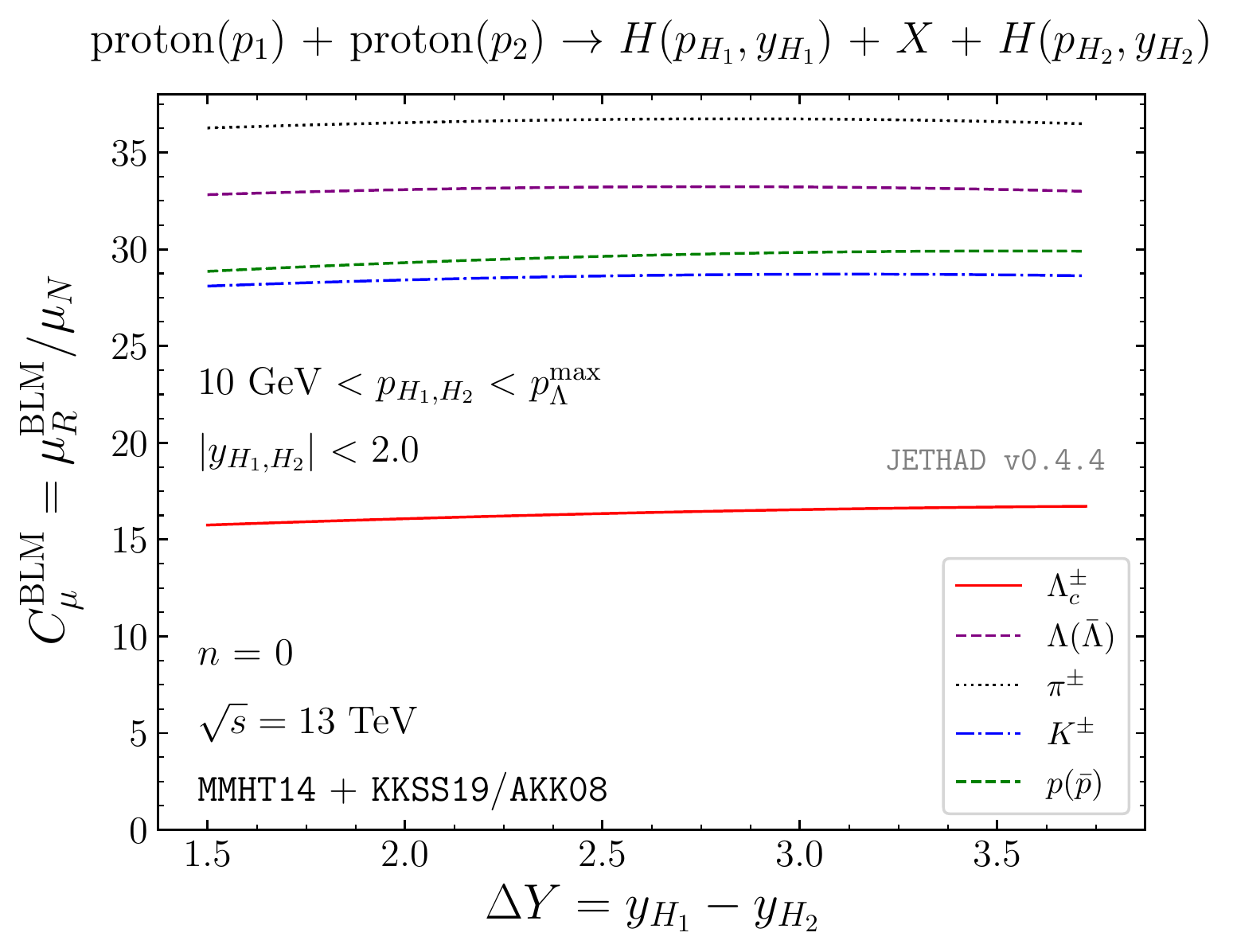}
   \includegraphics[scale=0.53,clip]{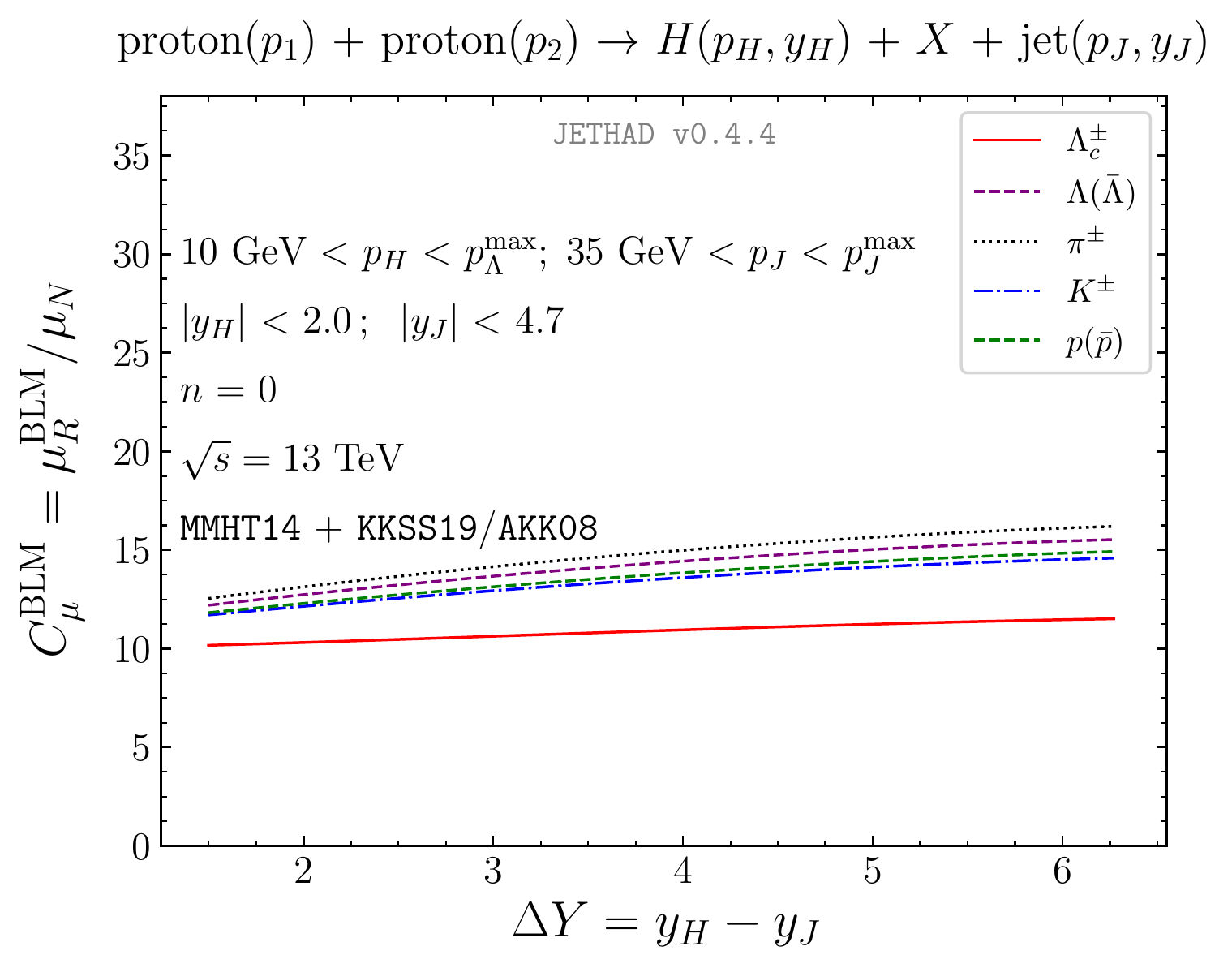}

   \includegraphics[scale=0.53,clip]{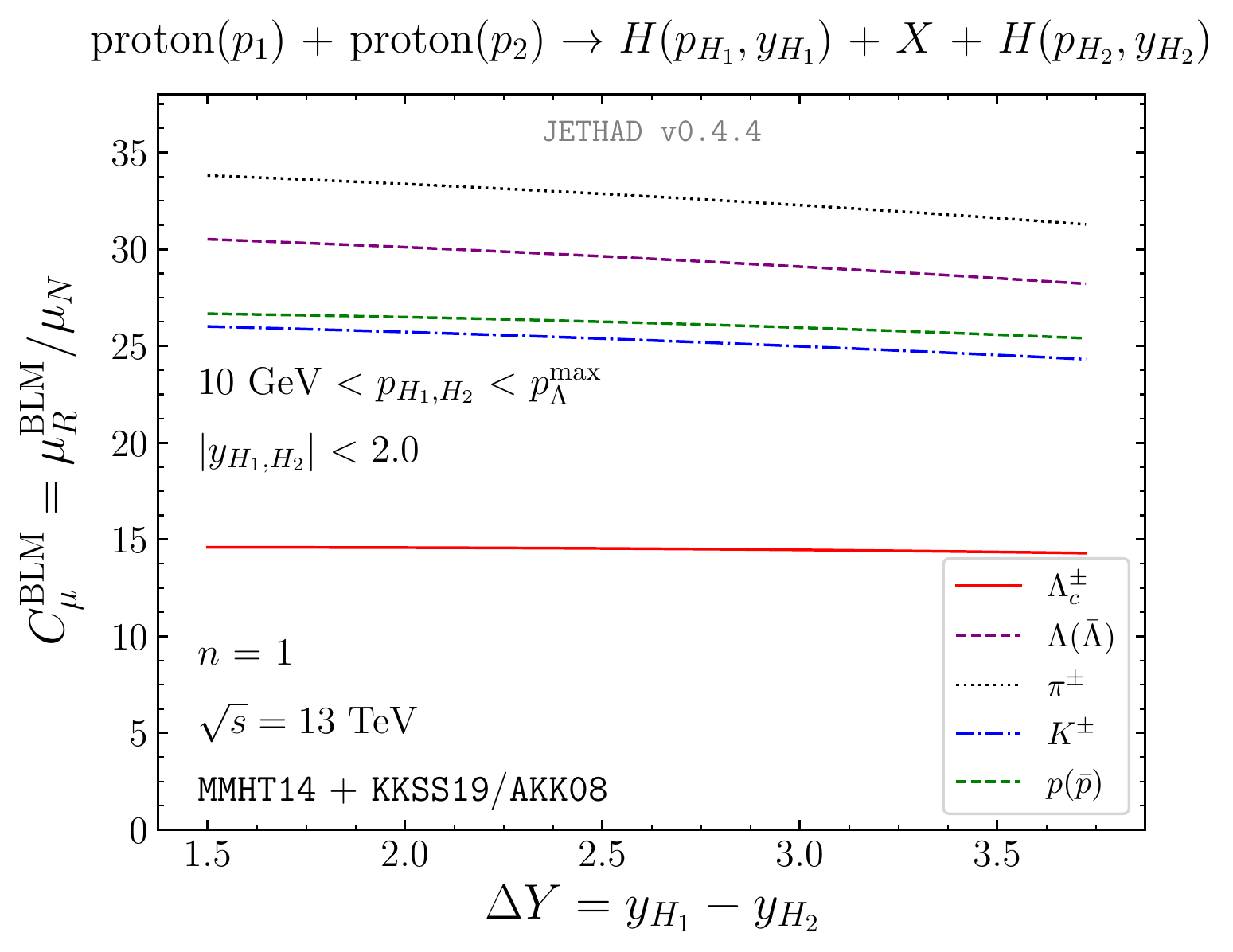}
   \includegraphics[scale=0.53,clip]{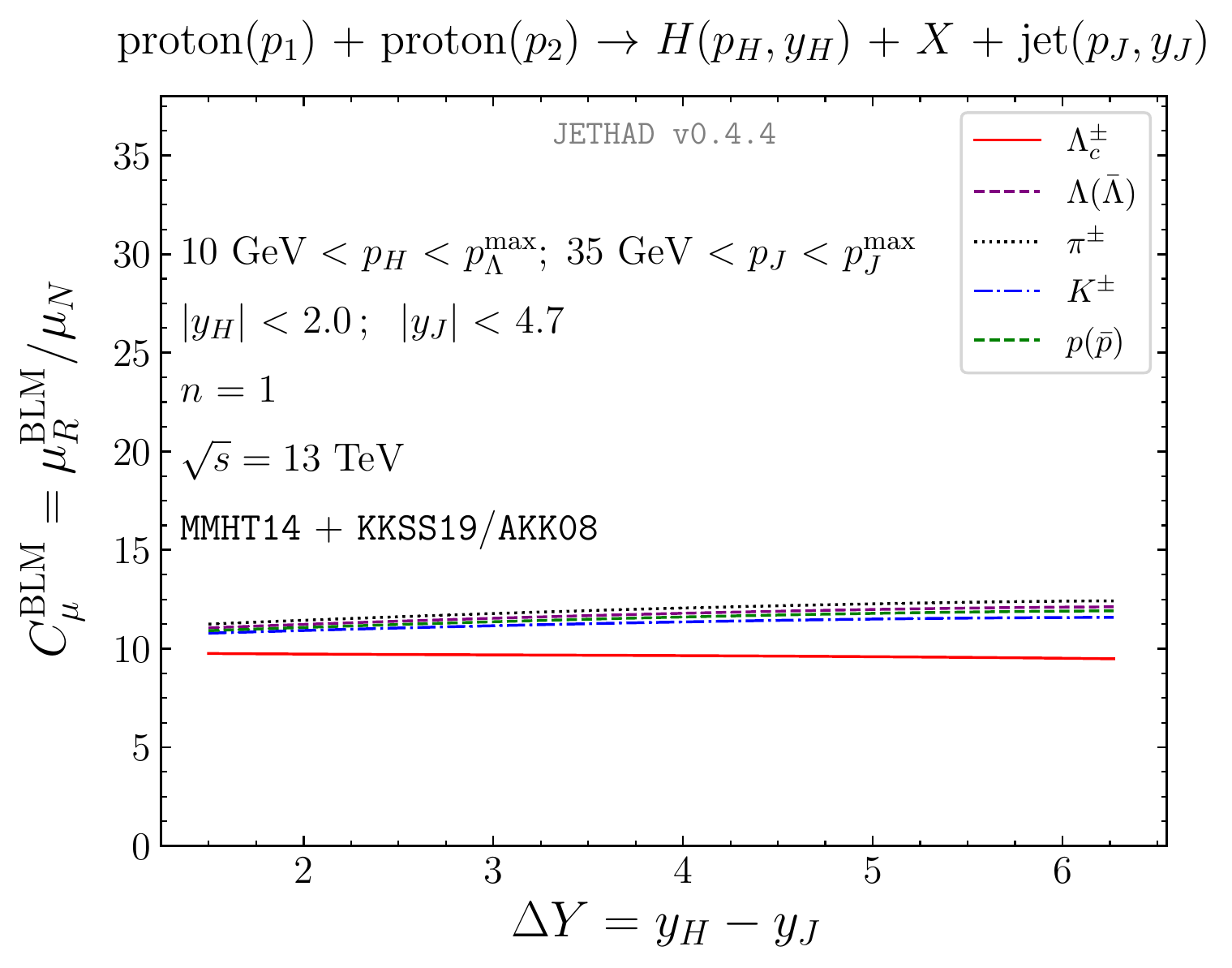}

   \includegraphics[scale=0.53,clip]{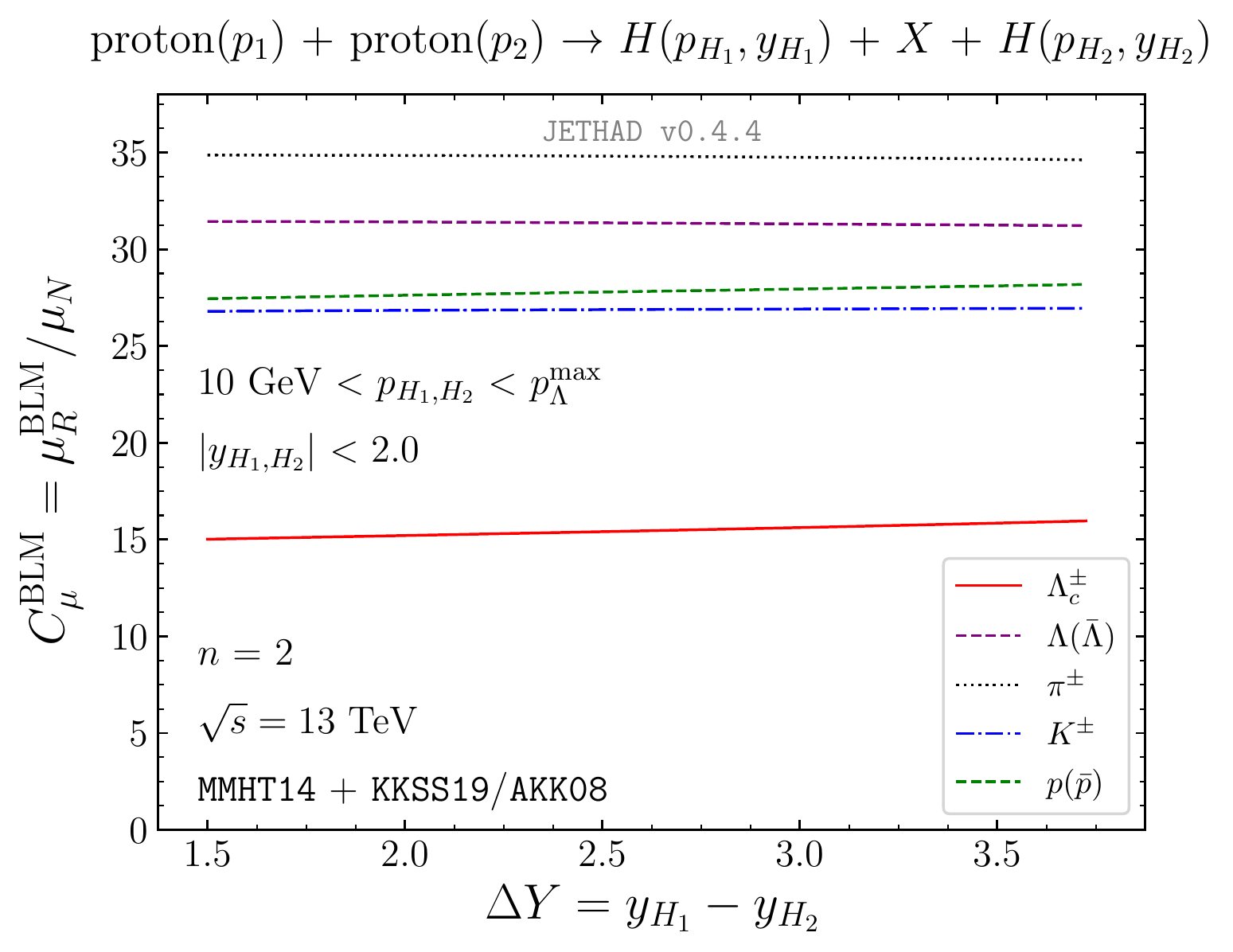}
   \includegraphics[scale=0.53,clip]{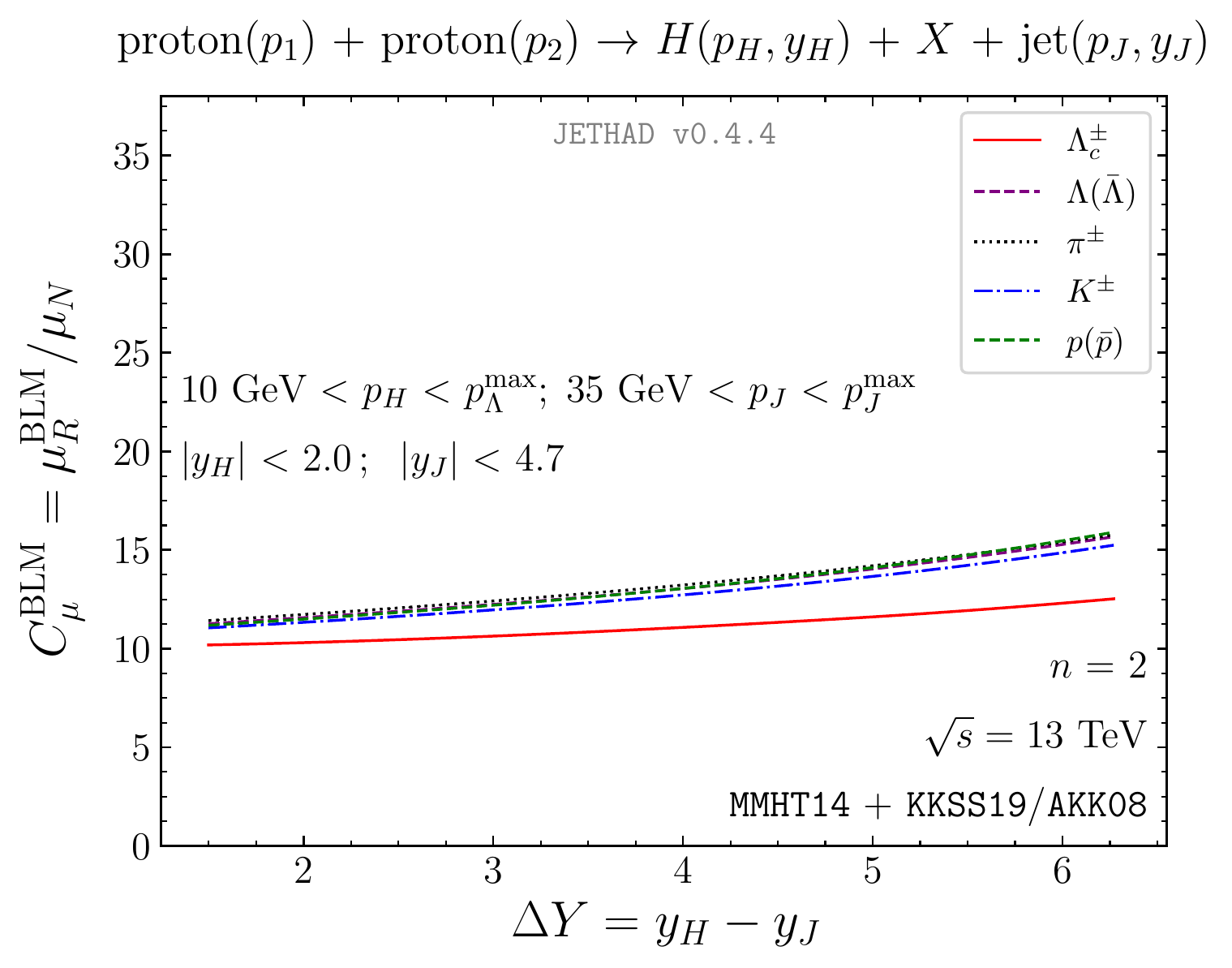}

\caption{BLM scales for the double $\Lambda_c$ (left) and the $\Lambda_c$ plus jet (right) production as functions of the rapidity separation, $\Delta Y$, for $n = 0, 1, 2$, and for $\sqrt{s} = 13$ TeV. Predictions for $\Lambda_c$ emissions are compared with configurations where other hadron species are tagged: $\Lambda(\bar{\Lambda})$, $\pi^\pm$, $K^\pm$, and $p(\bar{p})$. Text boxes inside panels show transverse-momentum and rapidity ranges.}
\label{fig:BLM_scales_HSA}
\end{figure}

As a preliminary analysis, we compare results for the BLM-scale parameter, $C_\mu^{\rm BLM}$, as a function of the rapidity interval, $\Delta Y$, for different species of emitted hadrons ($\Lambda$, $\pi$, $K$ and $p$). Notably, BLM scales for $\Lambda_c$ emissions are much lower (although still larger than natural ones) than the ones obtained when lighter-hadron species are detected. The effect is much more evident in the double $\Lambda_c$ channel (left panels of Fig.\tref{fig:BLM_scales_HSA}) with respect to the $\Lambda$ plus jet one (right panels of Fig.\tref{fig:BLM_scales_HSA}), this making us speculate that the production $\Lambda_c$ baryons could act as a \emph{stabilizer} of the BFKL series under higher-order corrections. Indeed, since applying the BLM method effectively translates into a rise of energy-scale values that reduces the weight of next-to-leading contributions, lower values of $C_\mu^{\rm BLM}$ indicate that the high-energy series was already (partially) stable, before applying BLM. A straightforward way to corroborate this statement consists in comparing predictions for observables of our interest both at natural scales and at BLM ones.

\begin{figure}[t]
\centering

   \includegraphics[scale=0.53,clip]{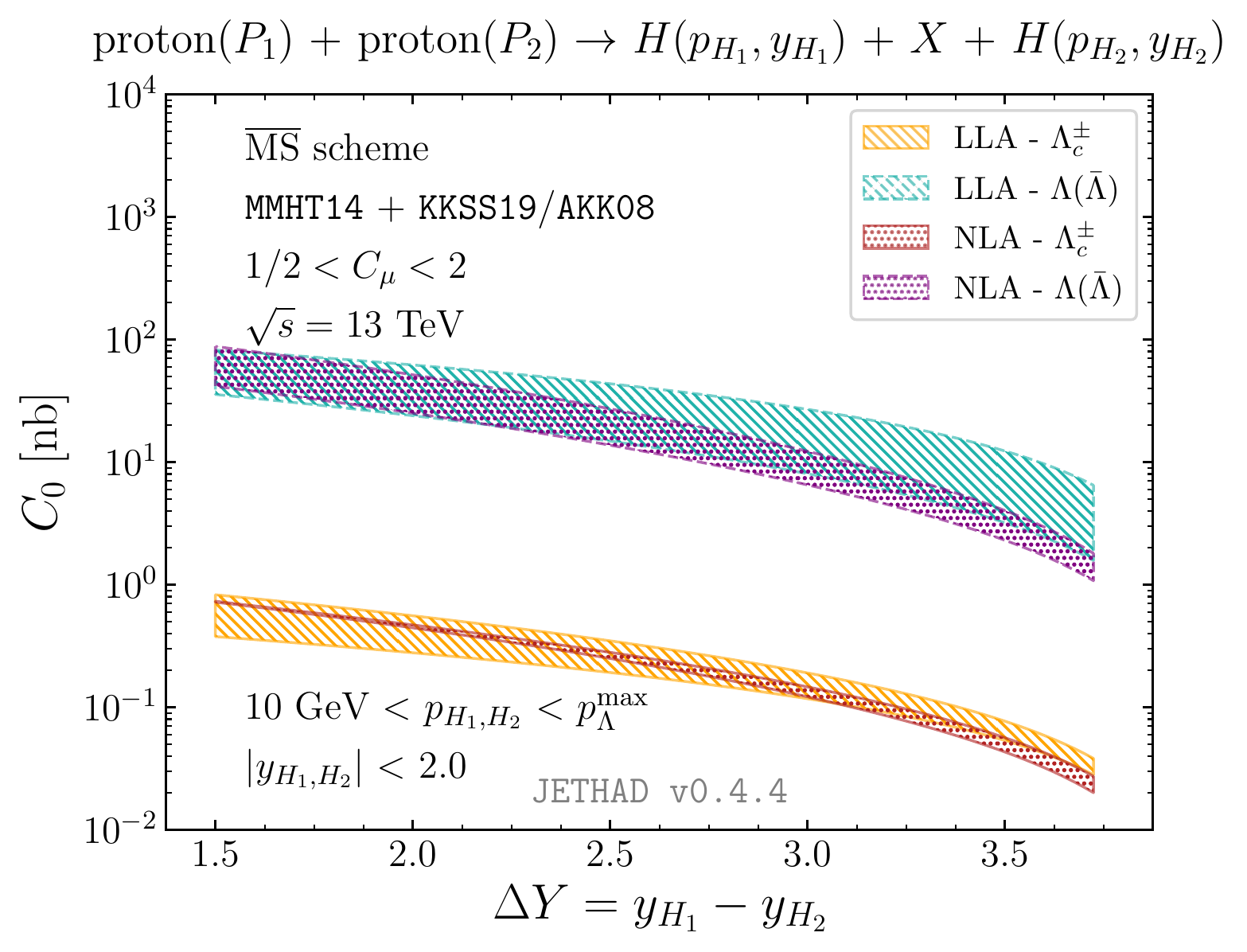}
   \includegraphics[scale=0.53,clip]{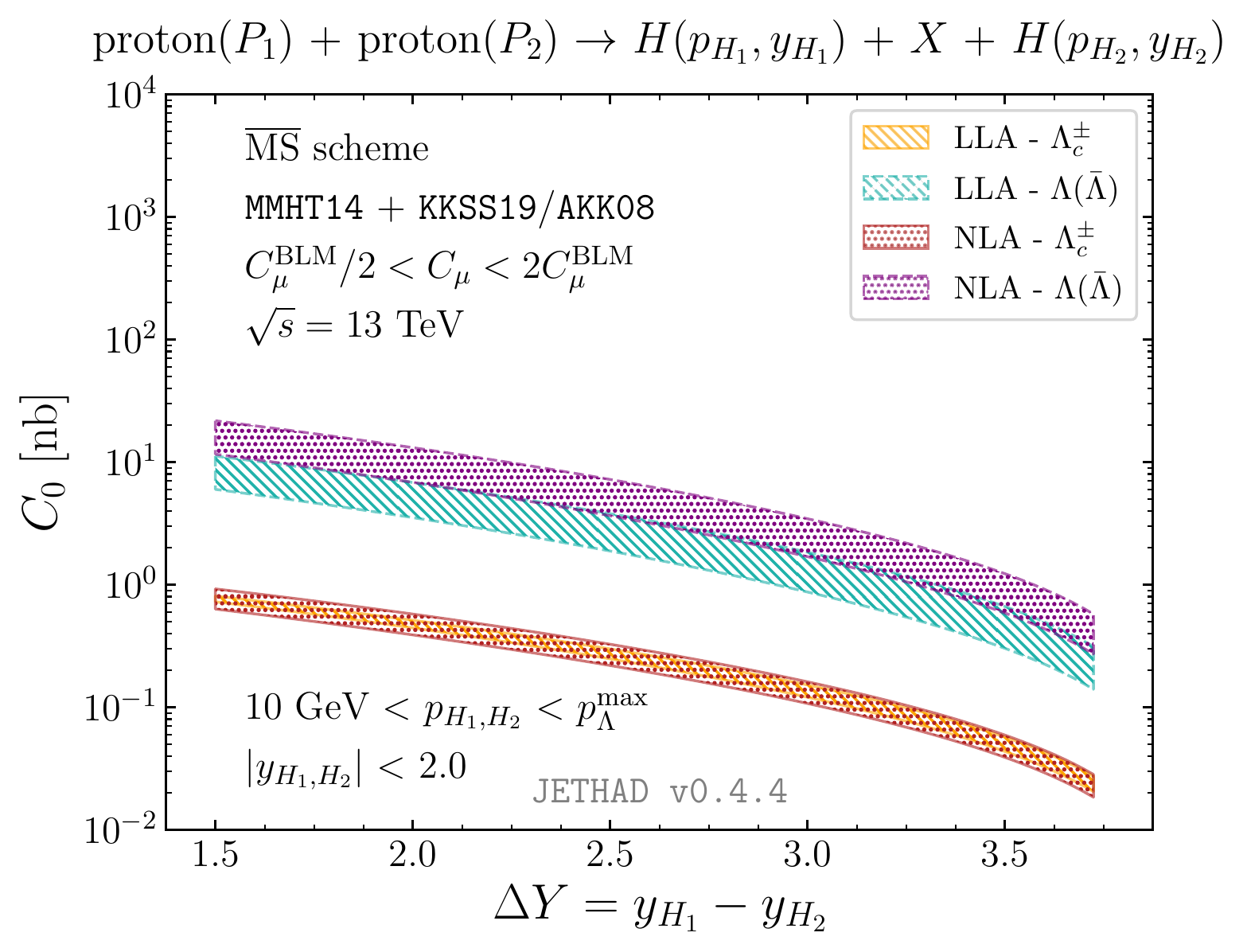}

   \includegraphics[scale=0.53,clip]{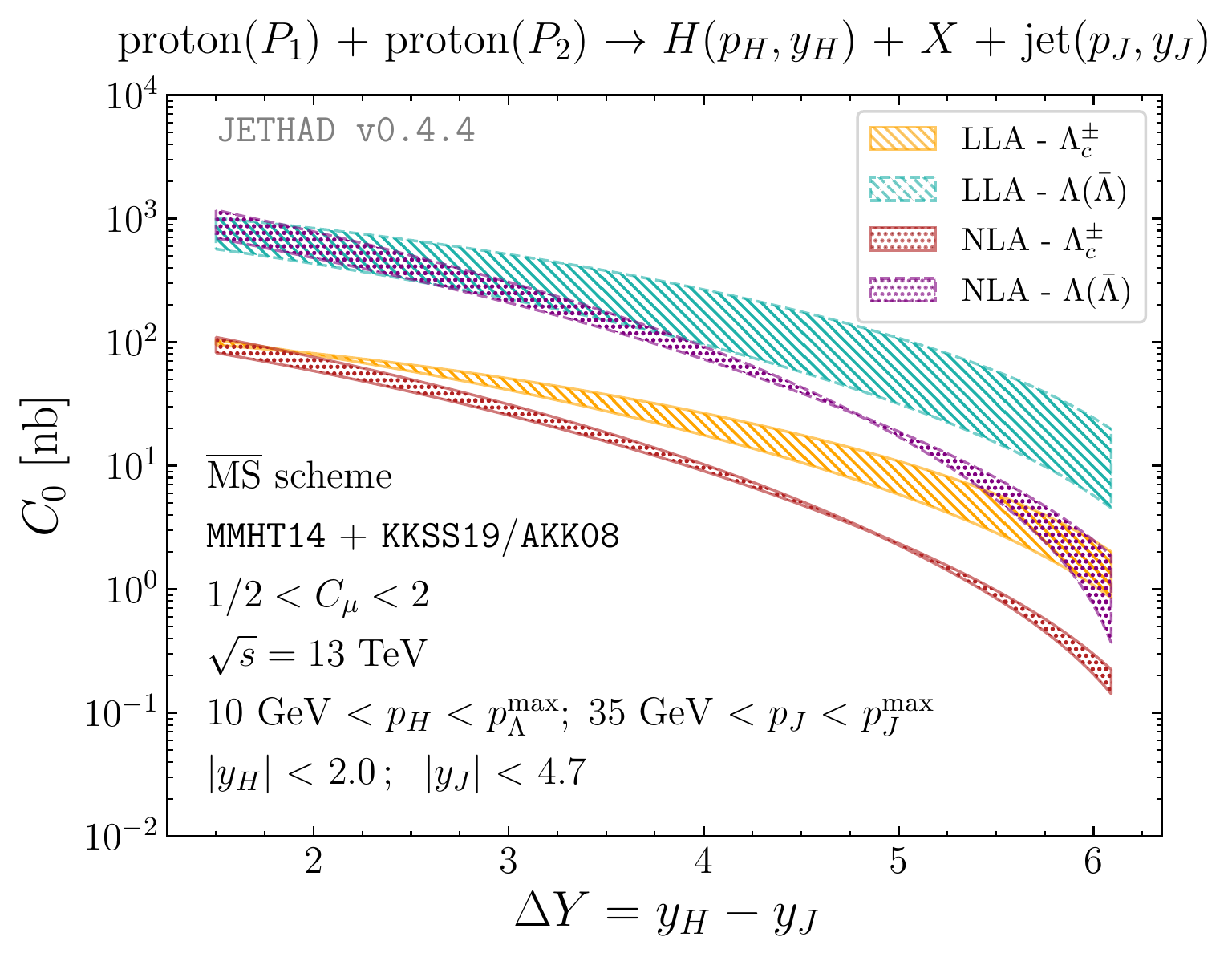}
   \includegraphics[scale=0.53,clip]{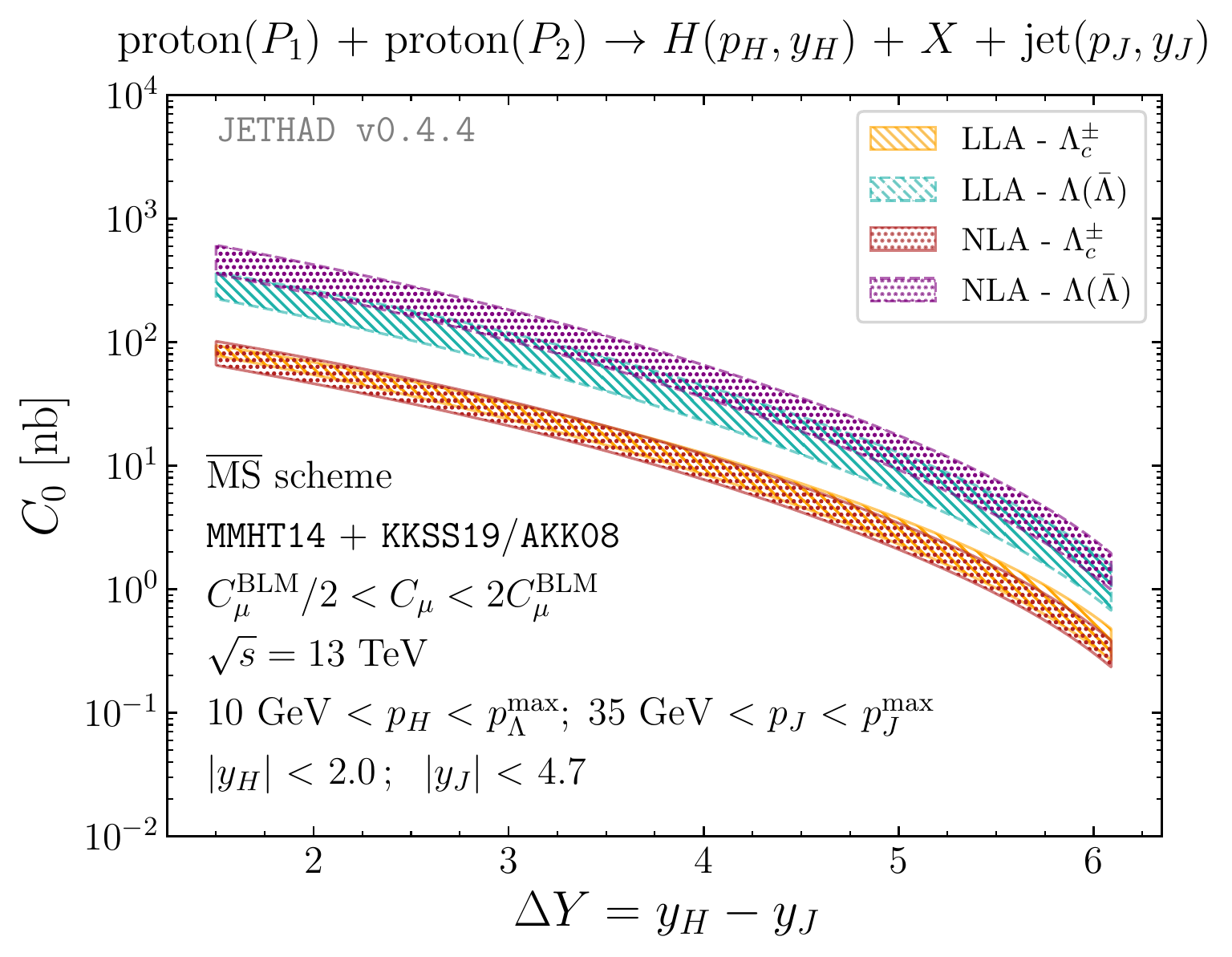}

\caption{Behavior of the $\varphi$-summed cross section, $C_0$, as a function of $\Delta Y$, in the double $\Lambda_c$ (upper) and in the $\Lambda_c$ plus jet channel (lower), at natural scales (left) and after BLM optimization (right), and for $\sqrt{s} = 13$ TeV. Error bands provide with the combined uncertainty coming from scale variation and numerical integrations. Predictions for $\Lambda_c$ emissions are compared with configurations where  $\Lambda$ hyperons are detected. Text boxes inside panels show transverse-momentum and rapidity ranges.}
\label{fig:C0_HSA}
\end{figure}

Upper panels of Fig.\tref{fig:C0_HSA} show the $\Delta Y$-dependence of the $\varphi$-summed cross section, $C_0$, in the double $\Lambda_c$ channel, together with corresponding predictions for detection of $\Lambda$ hyperons.
Here, two competing effects come into play. 
On the one hand, partonic cross sections grow with energy, as predicted by BFKL. On the other hand, collinear densities quench predictions when $\Delta Y$ increases. The net result is a downtrend with $\Delta Y$ of $C_0$ distributions.
We note that NLA bands are almost nested (except for very large values of $\Delta Y$) inside LLA ones and they are generally narrower in the $\Lambda_c$ case. This is a clear effect of a (partially) reached stability of the high-energy series, for both hadron emissions. However, while predictions for hyperons lose almost one order of magnitude when passing from natural scales to the expanded BLM ones (from left to right panel), results for $\Lambda_c$ baryons are much more stable, the NLA band becoming even wider in the BLM case.
The stability is partially lost when $\Lambda_c$ particle is accompanied by a jet, as shown in lower panels of Fig.\tref{fig:C0_HSA}. Here, LLA and NLA bands are almost disjoined at natural scales (left panel), while in the BLM case (right panel) they come closer to each other for hyperon plus jet, and almost entirely contained for $\Lambda_c$ plus jet. This study on $C_0$ clearly highlights how $\Lambda_c$ emissions allow for a stabilization of the resummed series, that cannot be obtained with lighter hadrons\footnote{For the sake of simplicity, we do not show here results for pions, kaons, and protons. It is known, however, that no such a stability can be reached via these channels (see Refs.\tcite{Celiberto:2017ptm,Bolognino:2018oth,Celiberto:2020wpk}).}, nor with the associated production of a light-flavored jet.
The detection of $\Lambda_c$ particles makes our distributions from one to two order of magnitude lower than the corresponding $\Lambda$ hyperon cases. This helps to dampen, from the experimental point of view, minimum-bias contaminations, in a more effective way when $\Lambda_c$ are tagged. At the same time, statistics is favorable, since values of $C_0$ for $\Lambda_c$ emissions are always larger than $10^{-2}$ nb.
We checked that the different behavior of our predictions between $\Lambda_c$ baryons and other hadrons is not artificially generated by the different lower cutoff for the momentum fraction in the native FF grids, $10^{-4}$ for {\tt KKSS19} versus $5\times10^{-2}$ for {\tt AKK08}. According to our numerical tests, both the $C_\mu^{\rm BLM}$ parameters and the $C_0$ distributions are left almost unchanged when the lower cutoff for {\tt KKSS19} is raised up to the {\tt AKK08} one. This feature was expected, since the major contribution to cross section is given by FF longitudinal fractions larger than $10^{-1}$ (see discussion Section in Ref.\tcite{Celiberto:2016hae}).

\begin{figure}[t]
\centering
\includegraphics[scale=0.53,clip]{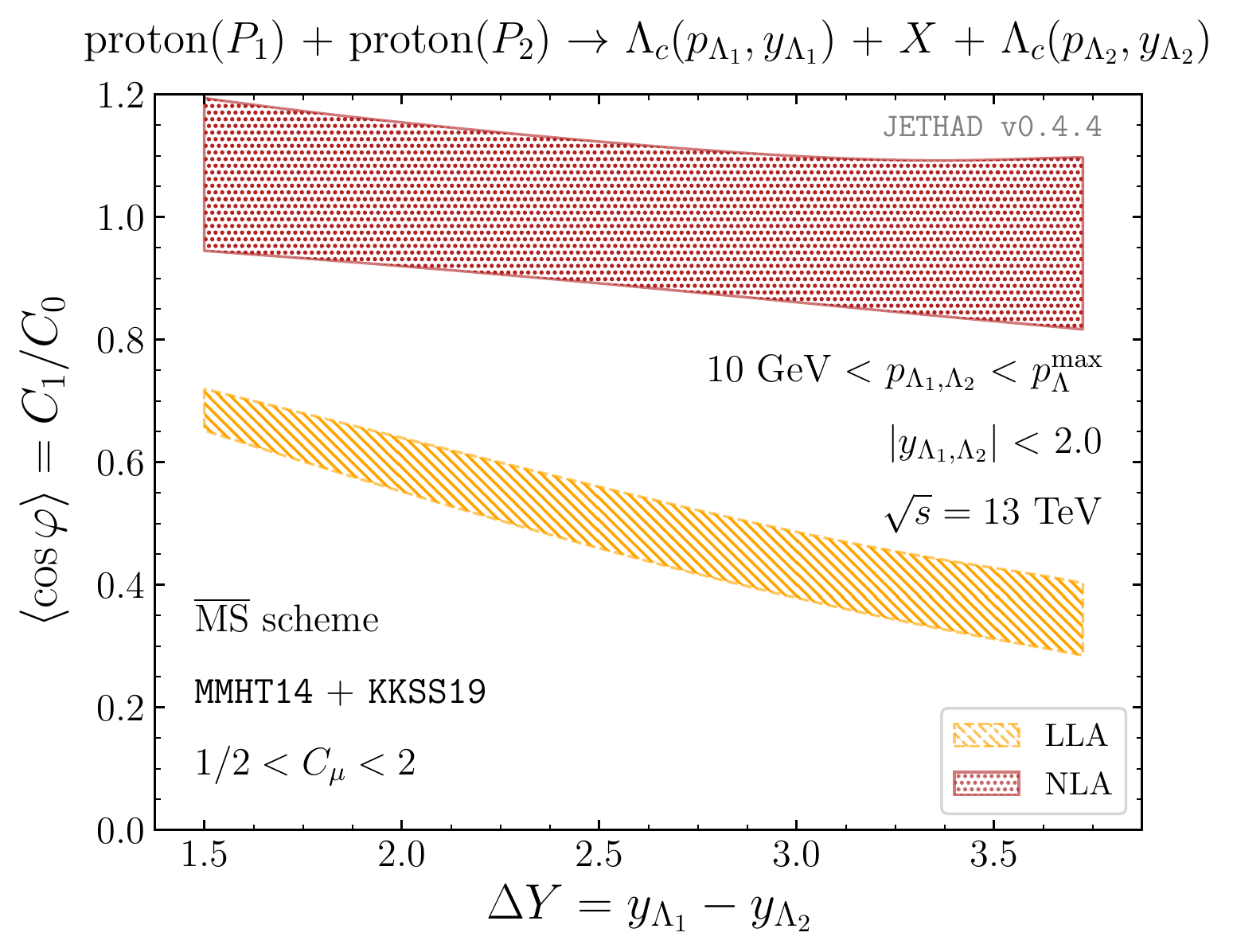}
\includegraphics[scale=0.53,clip]{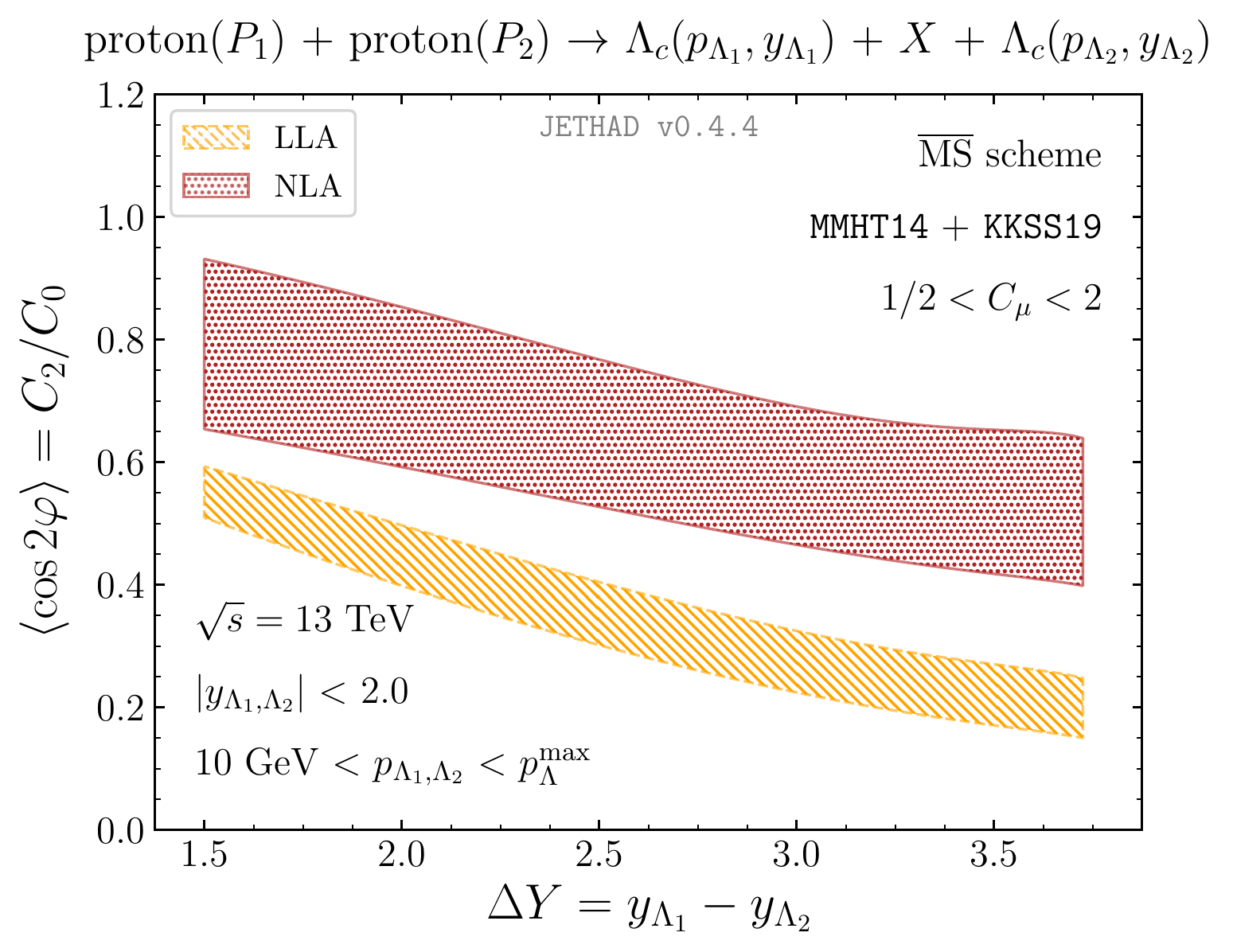}

\includegraphics[scale=0.53,clip]{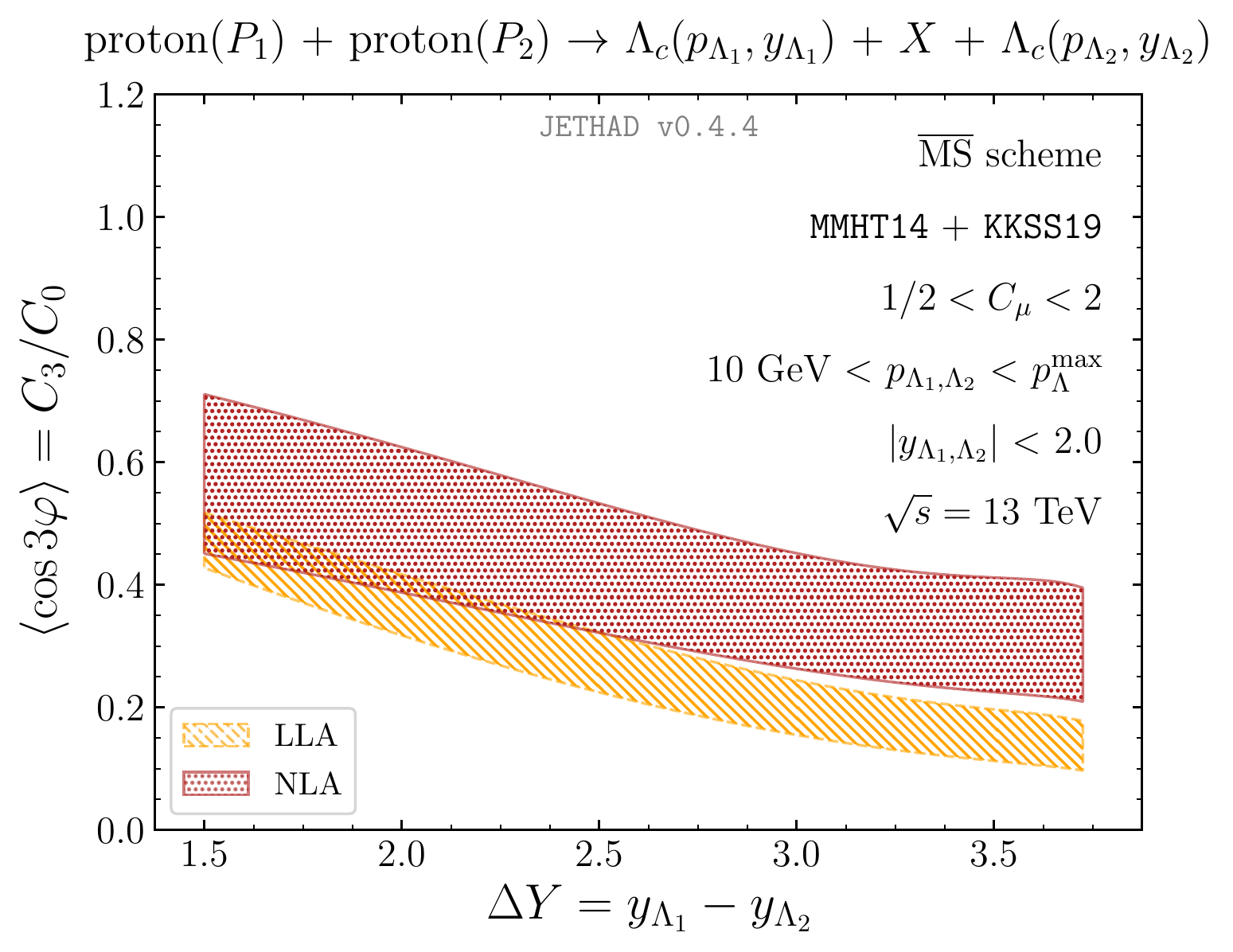}
\includegraphics[scale=0.53,clip]{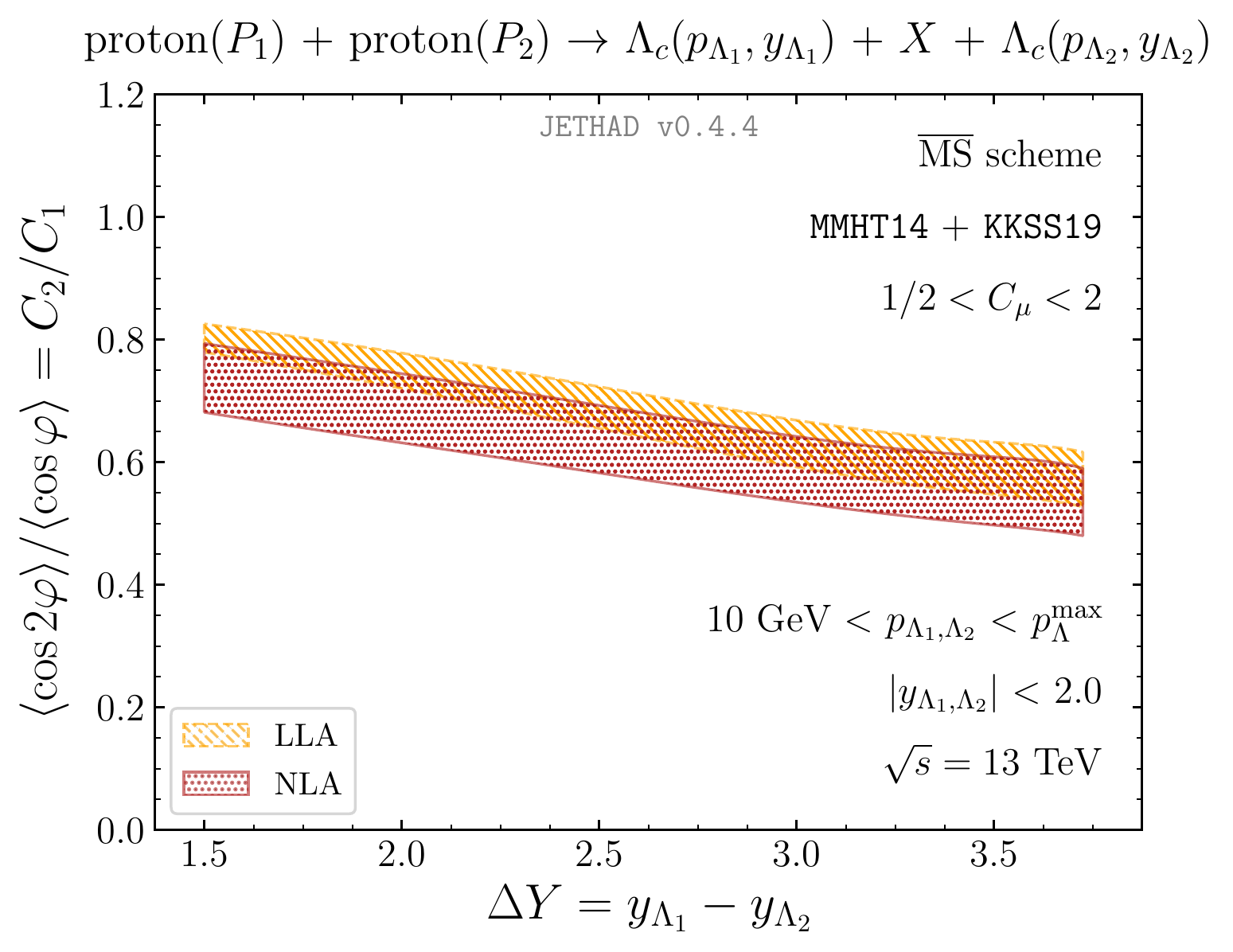}
\caption{Behavior of azimuthal-correlation moments, $R_{nm} \equiv C_{n}/C_{m}$, as functions of $\Delta Y$, in the double $\Lambda_c$ channel, at natural scales, and for $\sqrt{s} = 13$ TeV. Error bands provide with the combined uncertainty coming from scale variation and numerical integrations. Text boxes inside panels show transverse-momentum and rapidity ranges.}
\label{fig:Rnm_LL_NS}
\end{figure}

\begin{figure}[t]
\centering
\includegraphics[scale=0.53,clip]{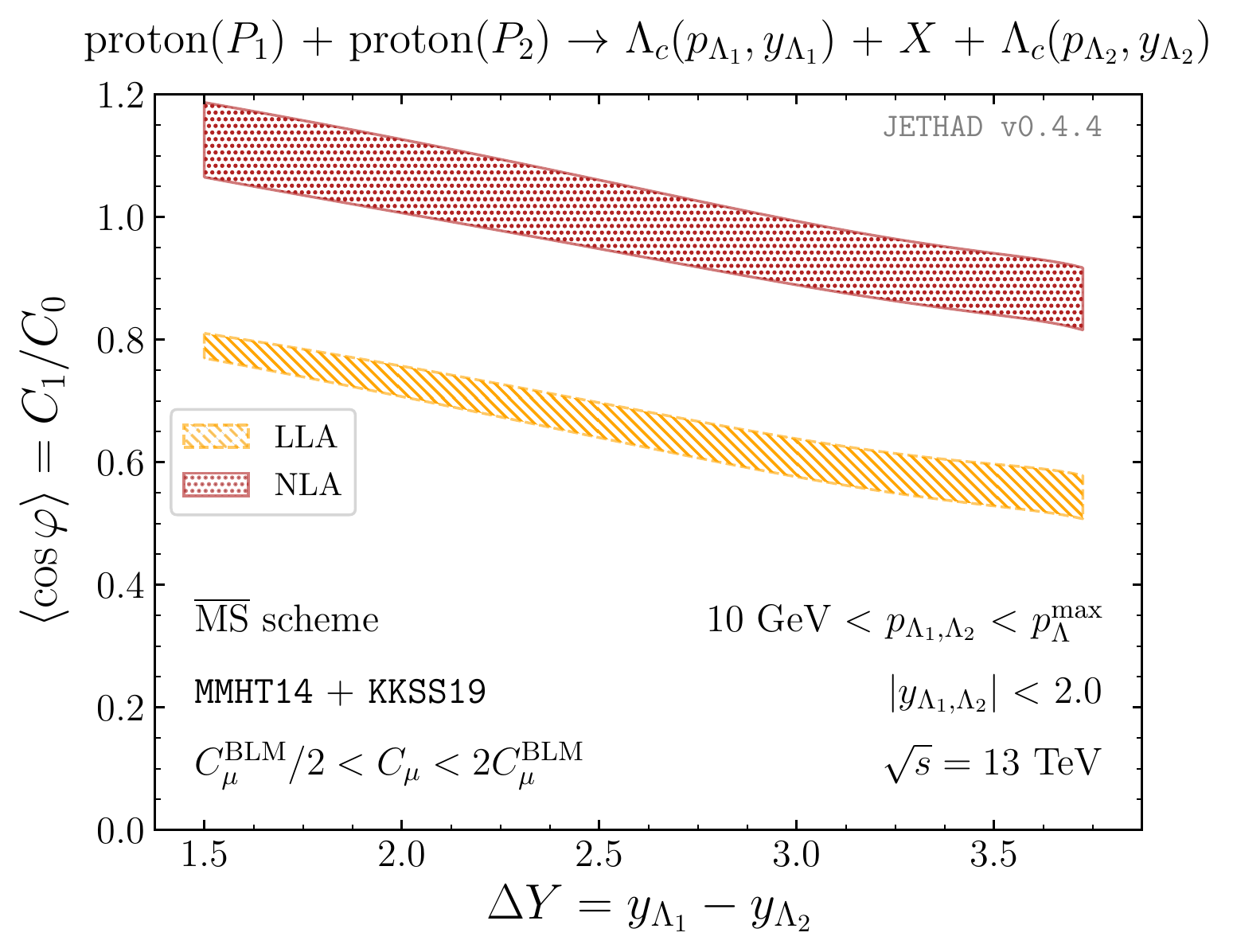}
\includegraphics[scale=0.53,clip]{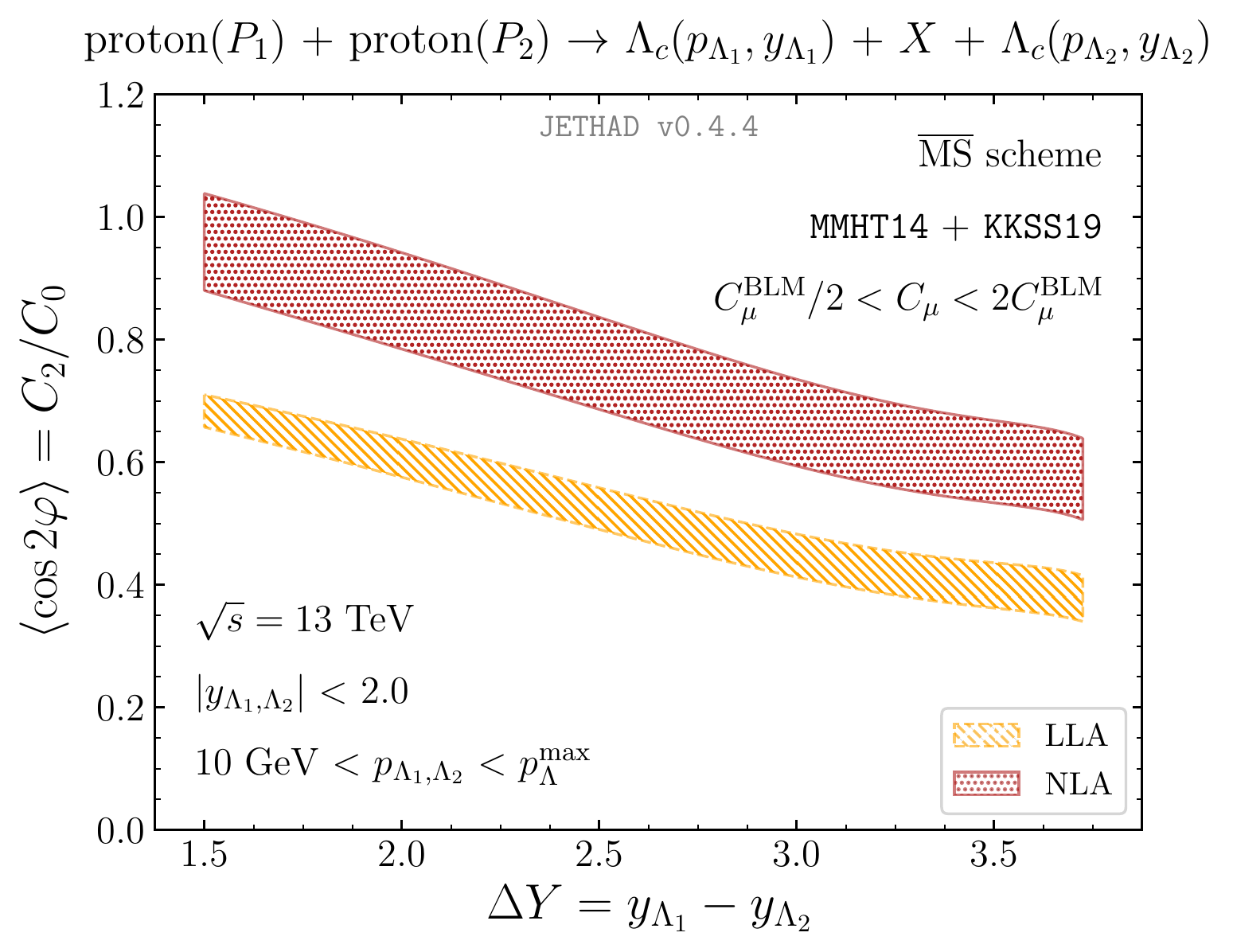}

\includegraphics[scale=0.53,clip]{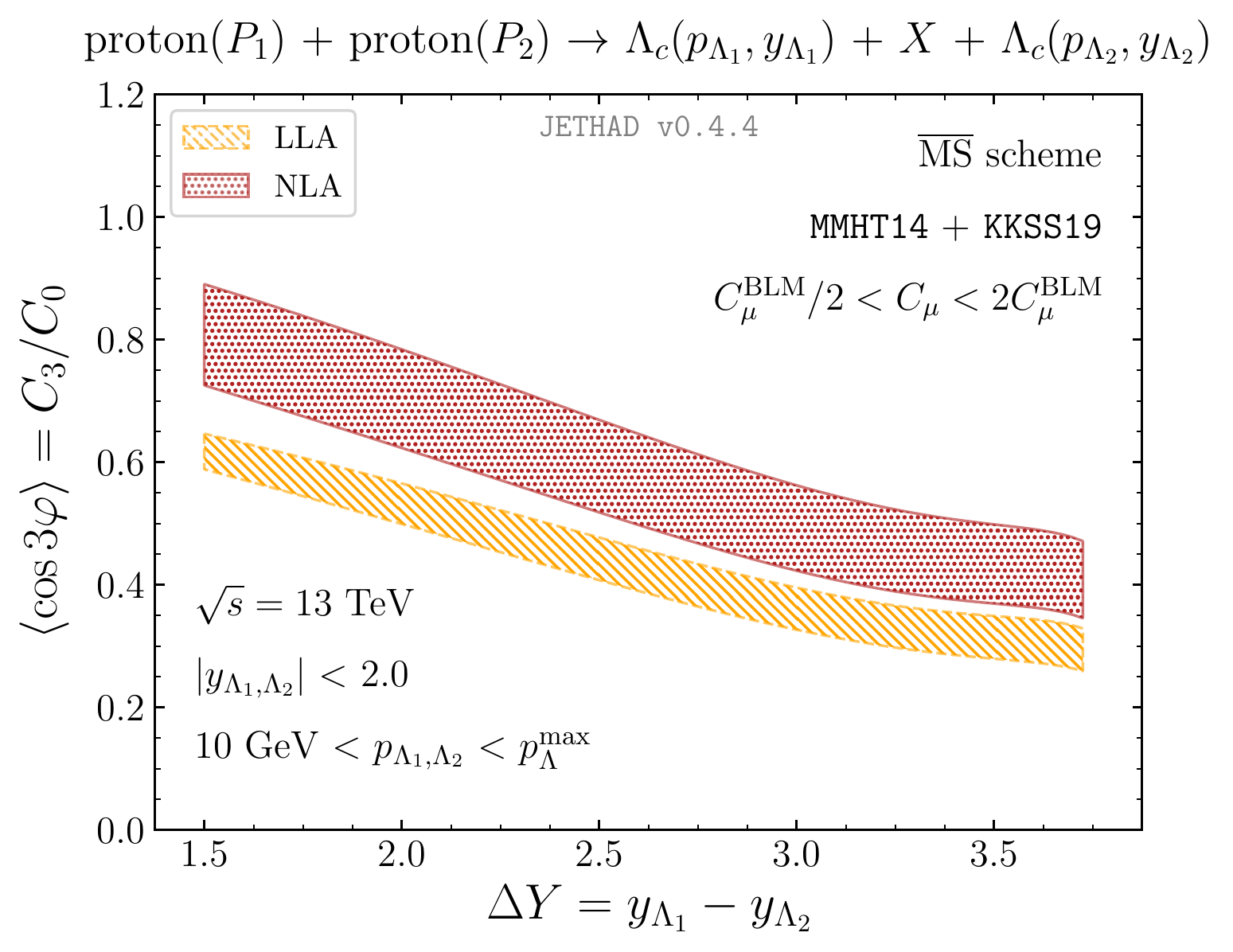}
\includegraphics[scale=0.53,clip]{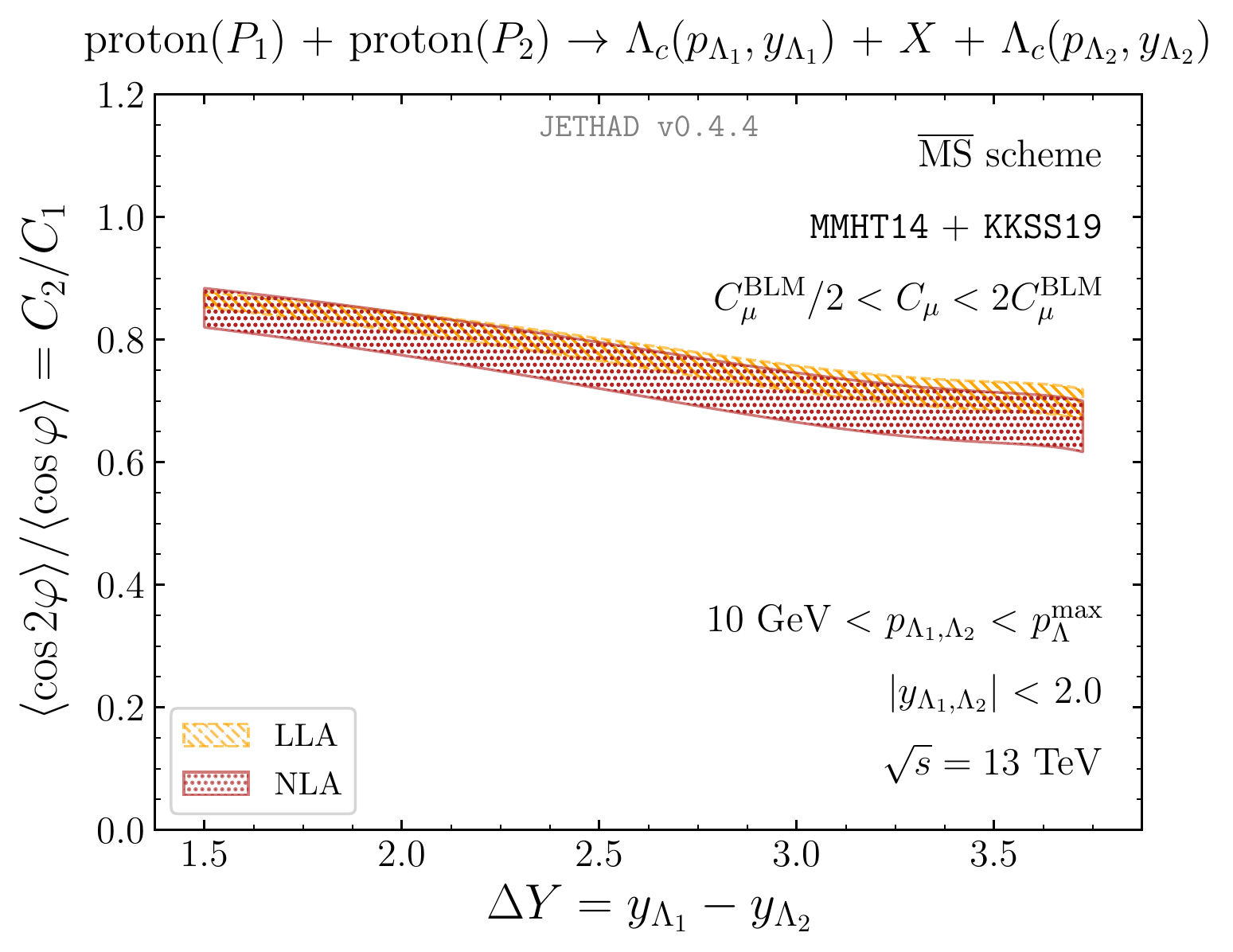}
\caption{Behavior of azimuthal-correlation moments, $R_{nm} \equiv C_{n}/C_{m}$, as functions of $\Delta Y$, in the double $\Lambda_c$ channel, at BLM scales, and for $\sqrt{s} = 13$ TeV. Error bands provide with the combined uncertainty coming from scale variation and numerical integrations. Text boxes inside panels show transverse-momentum and rapidity ranges.}
\label{fig:Rnm_LL_BLM}
\end{figure}

\begin{figure}[t]
\centering
\includegraphics[scale=0.53,clip]{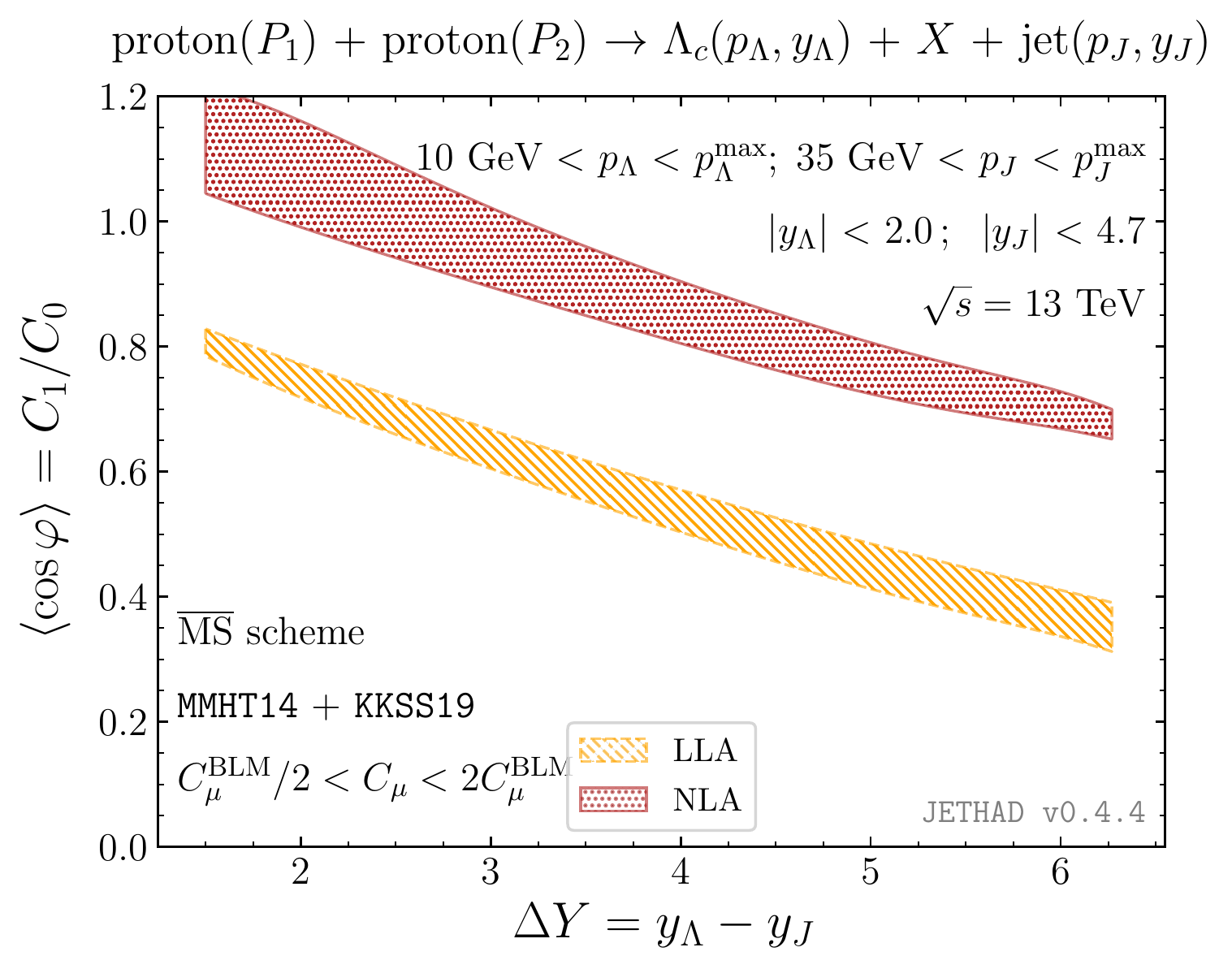}
\includegraphics[scale=0.53,clip]{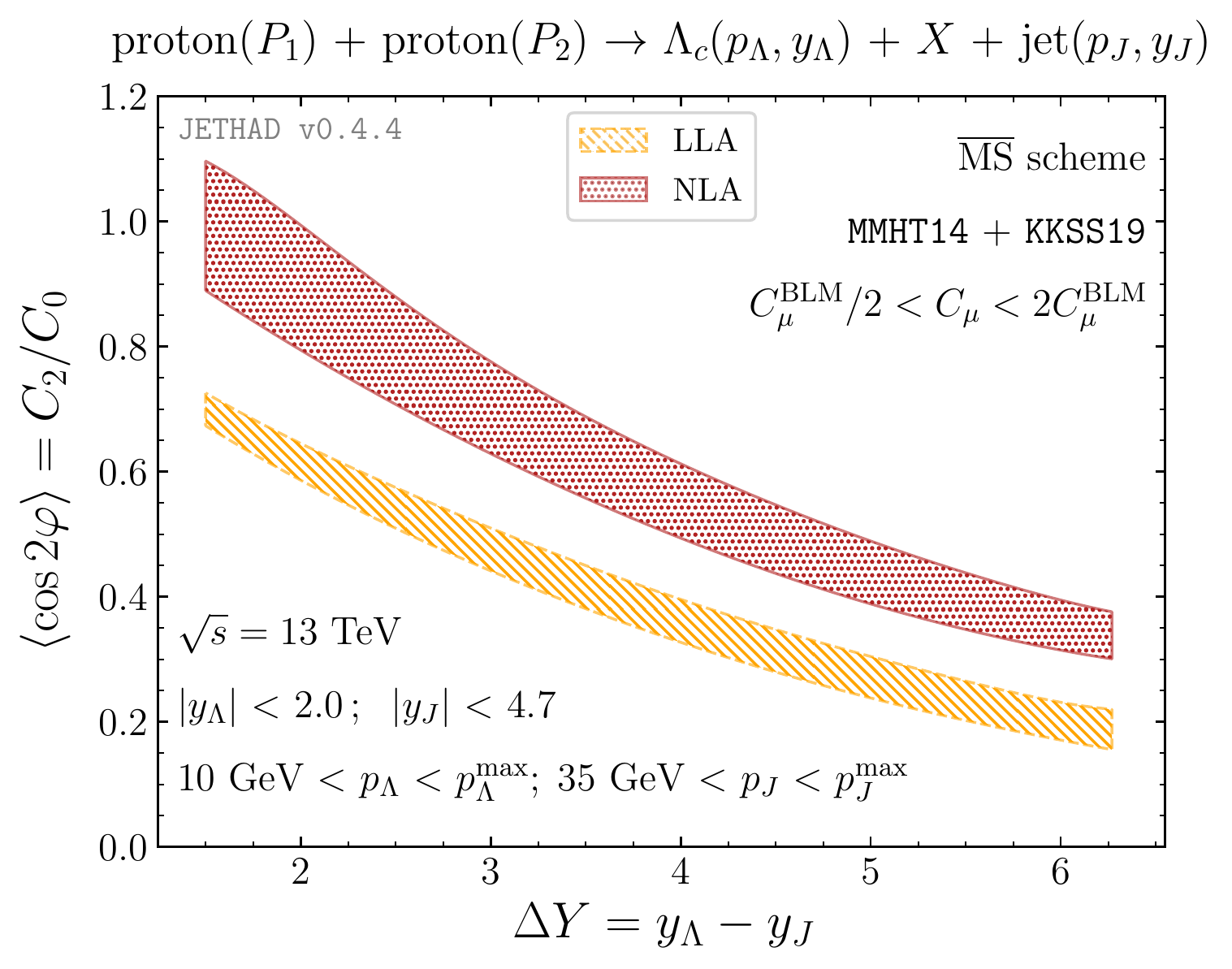}

\includegraphics[scale=0.53,clip]{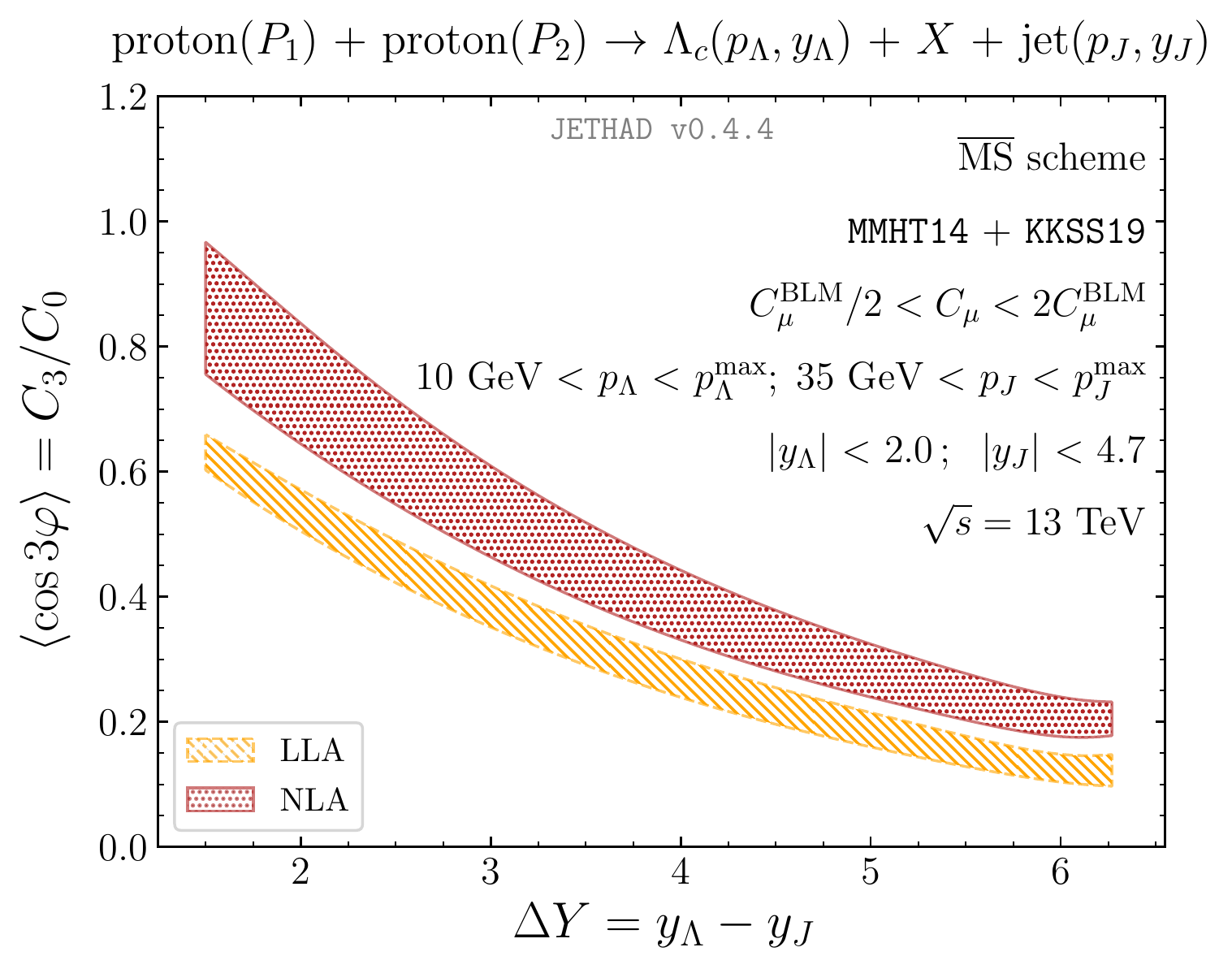}
\includegraphics[scale=0.53,clip]{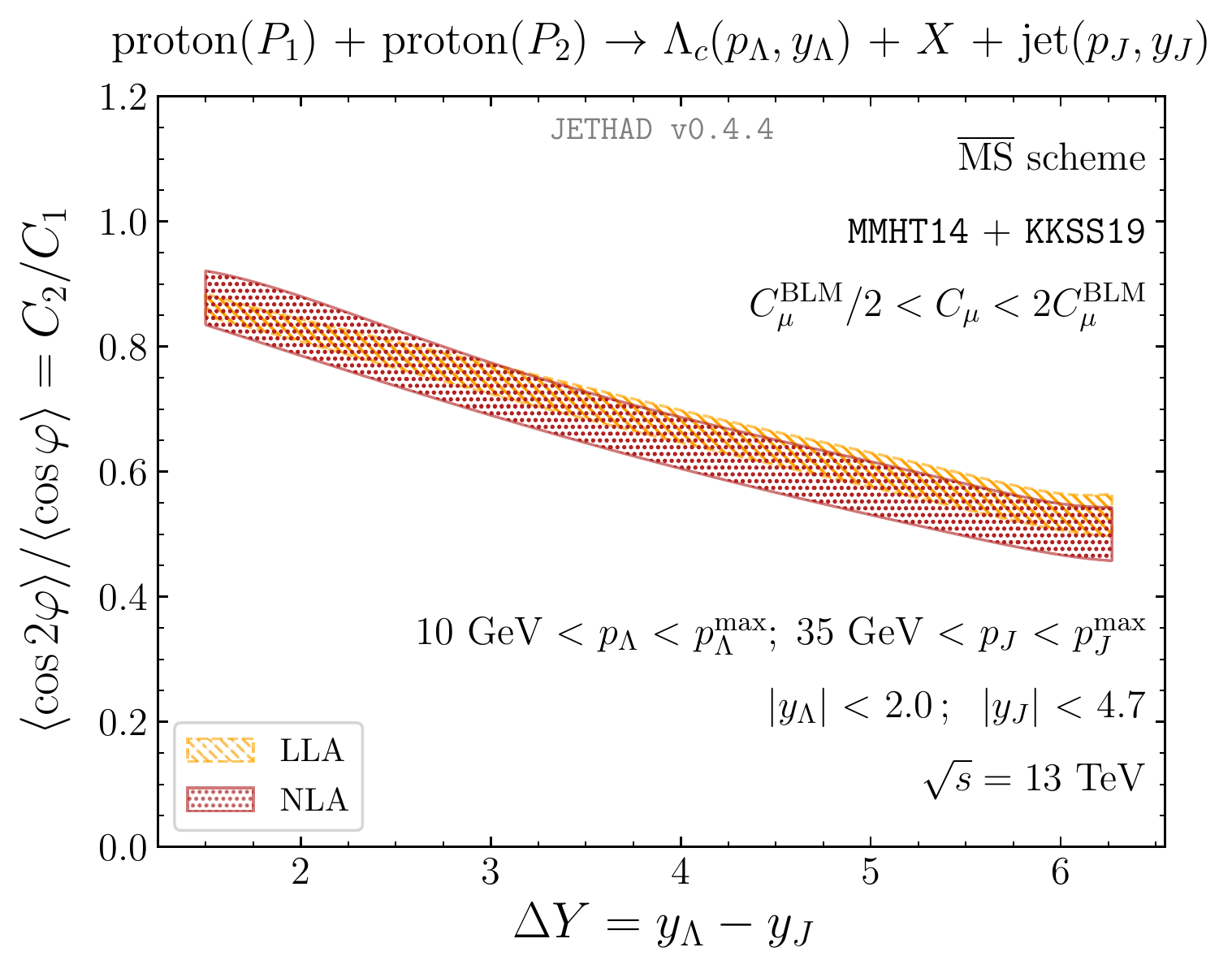}
\caption{Behavior of azimuthal-correlation moments, $R_{nm} \equiv C_{n}/C_{m}$, as functions of $\Delta Y$, in the $\Lambda_c$ plus jet channel, at BLM scales, and for $\sqrt{s} = 13$ TeV. Error bands provide with the combined uncertainty coming from scale variation and numerical integrations. Text boxes inside panels show transverse-momentum and rapidity ranges.}
\label{fig:Rnm_LJ_BLM}
\end{figure}

Predictions for $R_{nm}$ azimuthal ratios in the double $\Lambda_c$ channel at natural and at BLM scales are presented in Figs.\tref{fig:Rnm_LL_NS} and \tref{fig:Rnm_LL_BLM}, respectively.
The downtrend of all these ratios when $\Delta Y$ grows is a well-know signal of the onset of high-energy dynamics.
Larger rapidity distances heighten the weight of undetected gluons, thus leading to a decorrelation pattern in the azimuthal plane, which is more pronounced in pure LLA series. At variance with the $C_0$ case, here we observe a reversed situation, where NLA BLM-optimized results exhibit a narrower uncertainty bands and are closer to the corresponding LLA ones, with respect to what happens at natural scales.
This dichotomy is much more emphasized in the $\Lambda_c$ plus jet channel. In particular, instabilities rising at natural scales are strong enough (although being milder than the ones observed in the Mueller--Navelet dijet channel) to prevent any realistic analysis. Therefore, we show only the behavior of $R_{nm}$ ratios after BLM optimization (Fig.\tref{fig:Rnm_LJ_BLM}), whose patterns are in line with corresponding predictions for the double $\Lambda_c$ production. The $R_{21}$ ratio exhibits a fair stability under NLA corrections for both scale choices and both final-state channels (see lower left panels of Figs.\tref{fig:Rnm_LL_NS}-\ref{fig:Rnm_LJ_BLM}). In the presented plots, the value of the $R_{10}$ moment exceeds one in the small-$\Delta Y$ region. This unphysical effect is fairly explained by the fact that contributions which are power-suppressed in energy and are not included in our BFKL treatment start to become relevant in those kinematic ranges, thus worsening the accuracy of our predictions.

\begin{figure}[t]
\centering
\includegraphics[scale=0.53,clip]{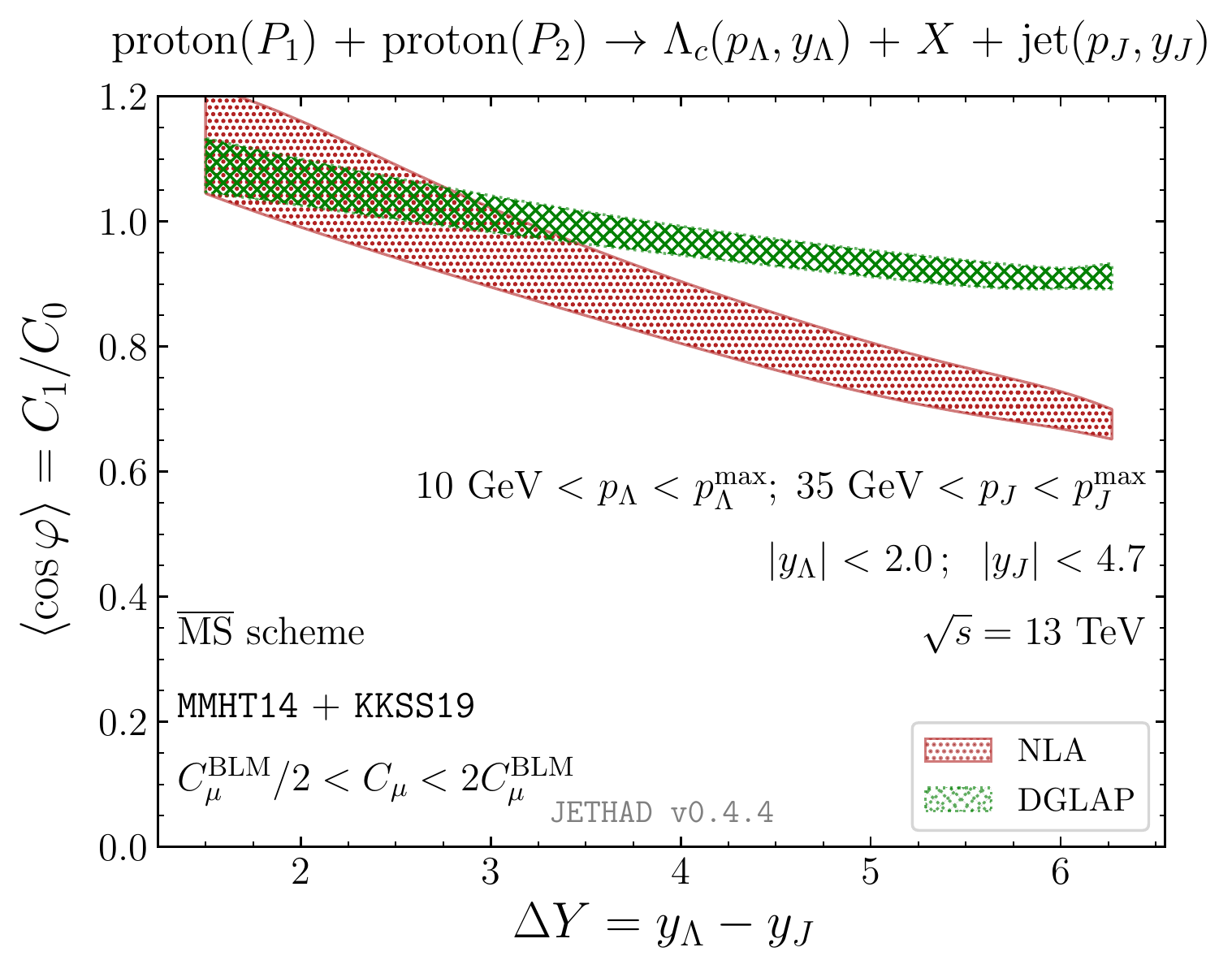}
\includegraphics[scale=0.53,clip]{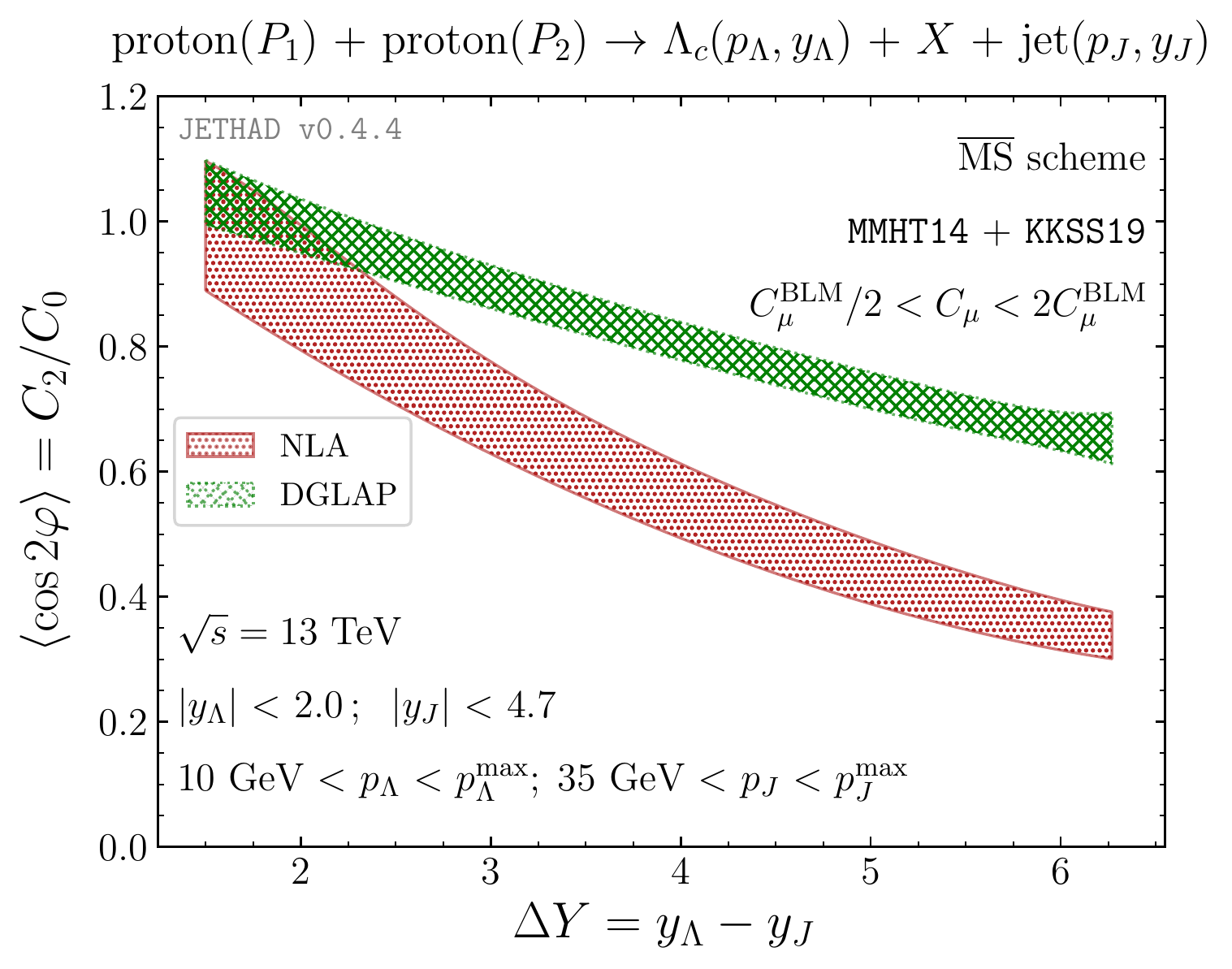}

\caption{BFKL-versus-DGLAP comparison for $R_{10} \equiv \langle \cos \varphi \rangle$ and $R_{20} \equiv \langle \cos 2 \varphi \rangle$ azimuthal-correlation moments, as functions of $\Delta Y$, in the $\Lambda_c$ plus jet channel, at BLM scales, and for $\sqrt{s} = 13$ TeV. Error bands provide with the combined uncertainty coming from scale variation and numerical integrations. Text boxes inside panels show transverse-momentum and rapidity ranges.}
\label{fig:Rnm_LJ_BLM_DGLAP}
\end{figure}

We complete our analysis by comparing NLA predictions for $\Lambda_c$ plus jet final-state configurations with the corresponding ones calculated in our high-energy DGLAP limit. The $\Delta Y$ dependence of $R_{10}$ and $R_{20}$ azimuthal correlations with BLM scale optimization is presented in Fig.\tref{fig:Rnm_LJ_BLM_DGLAP}. As expected, the distance between BFKL and DGLAP, already marked at low $\Delta Y$, becomes sharper and sharper when the rapidity interval grows. This phenomenon, already observed in the dijet\tcite{Celiberto:2015yba} as well as in the hadron plus jet channel\tcite{Celiberto:2020wpk,Celiberto:2020rxb}, is easily explained. At variance with the BFKL case, the limited number of inclusive gluon emissions due to the truncation of the perturbative series make the two final-state objects almost fully correlated (namely \emph{quasi} back-to-back produced) independently of the value of $\Delta Y$. The genuinely asymmetric kinematic configuration provided by the $\Lambda_c$ plus jet reaction suppresses the Born contribution, thus magnifying the distance between BFKL and DGLAP.

The overall outcome is that the inclusive detection of $\Lambda_c$ of baryons in semi-hard reactions allows for a stabilization of the high-energy resummation under higher-order corrections. A similar effect has been already observed in processes involving the production of massive bosons, such as the recently proposed Higgs plus jet channel\tcite{Celiberto:2020tmb}. However, while in that case the large energy scales provided by the Higgs transverse mass act as ``natural" stabilizers for the BFKL series, here it comes as an intrinsic feature of the $\Lambda_c$ production. Moreover, at variance with the Higgs plus jet case, where the formal description relies on a partial NLA treatment, here the stabilizing effect is manifest in the full NLA BFKL.

\subsection{More on $\Lambda_c$ fragmentation}
\label{more}

\begin{figure}[t]
\centering
\includegraphics[scale=0.53,clip]{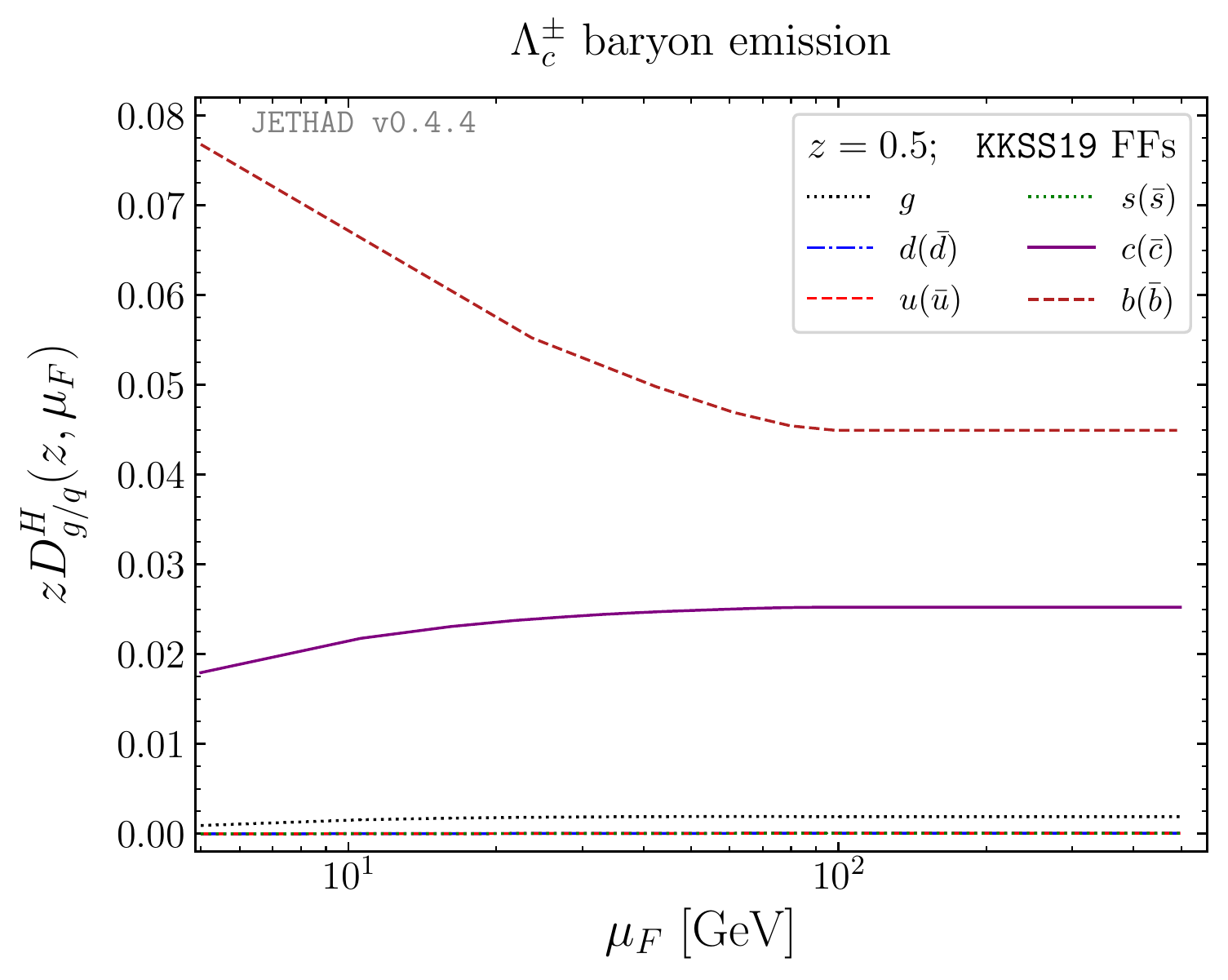}
\includegraphics[scale=0.53,clip]{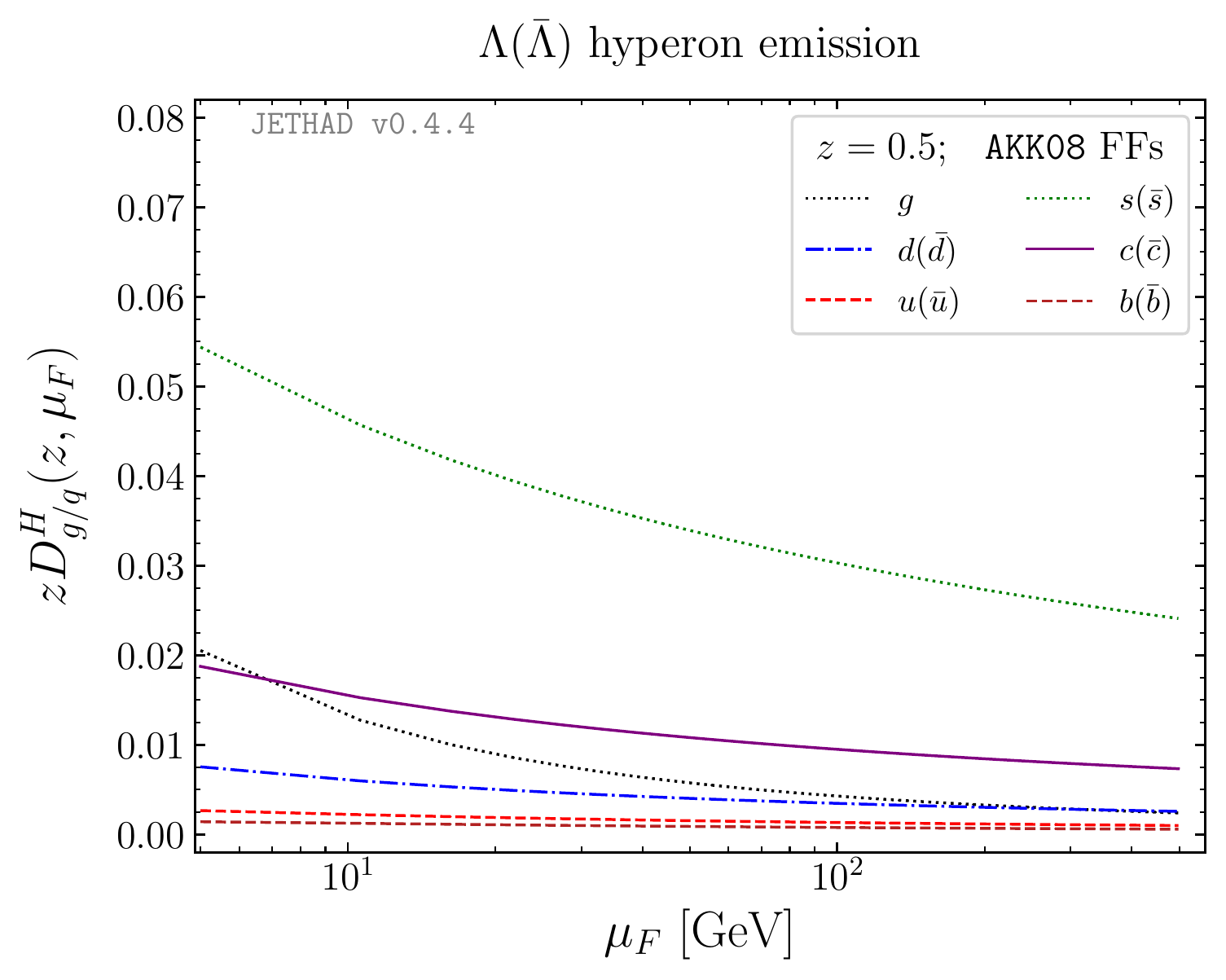}

\hspace{0.1cm} \includegraphics[scale=0.53,clip]{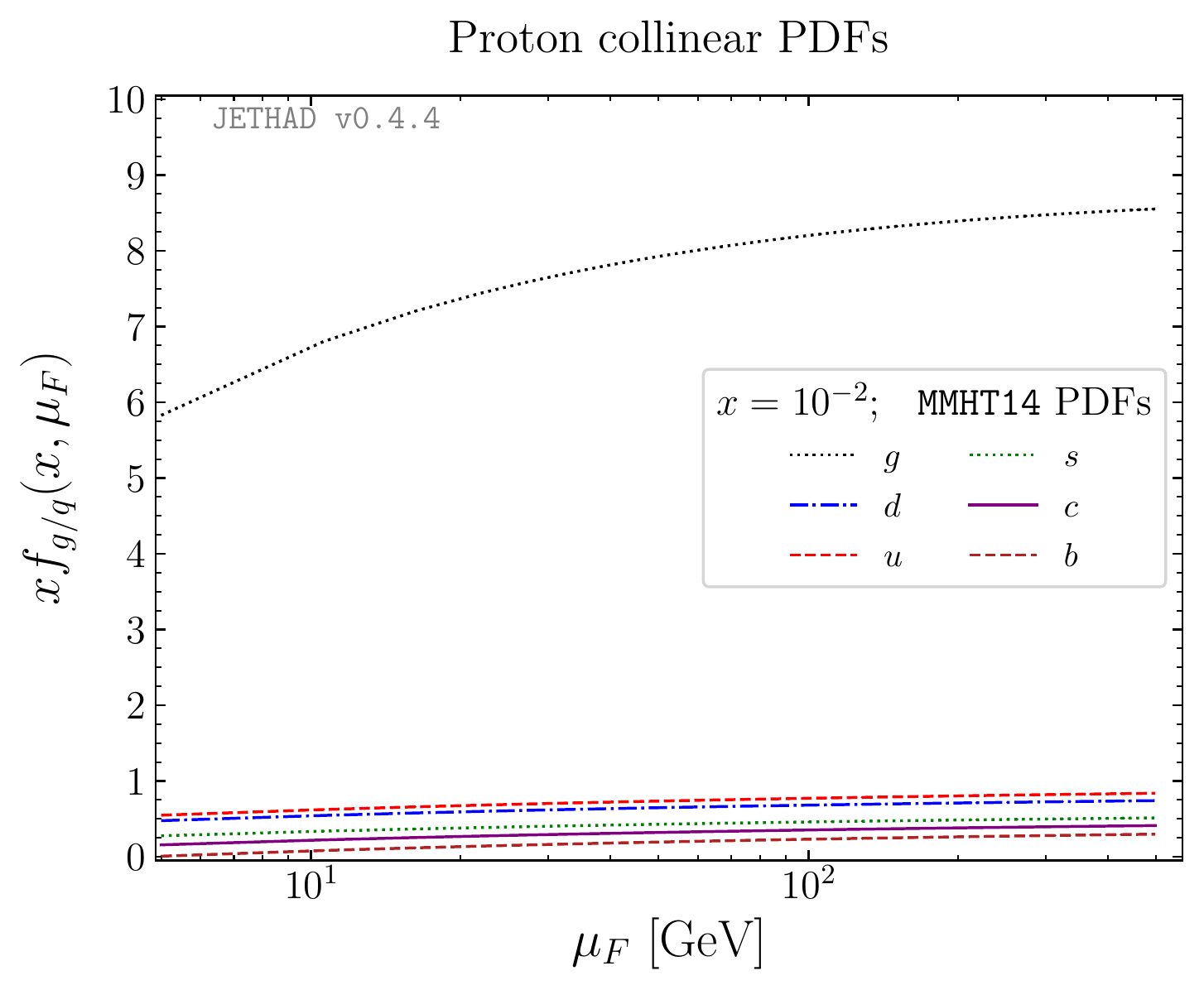}
\hspace{0.05cm} \includegraphics[scale=0.534,clip]{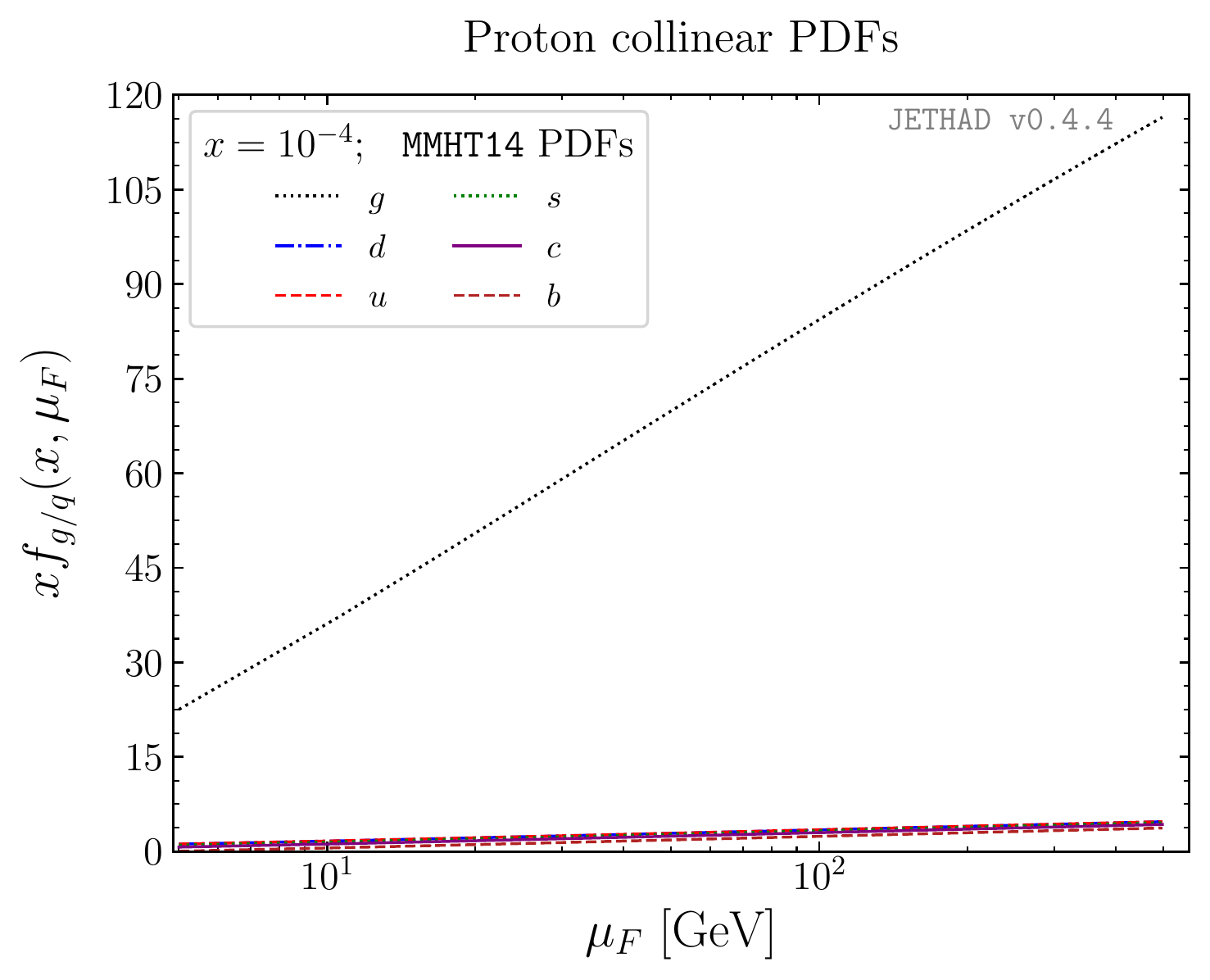}

\caption{Upper panels: energy-scale dependence of $\Lambda_c^{\pm}$ {\tt KKSS19} (left) and $\Lambda (\bar \Lambda)$ {\tt AKK08} (right) NLO FFs for $z = 5 \times 10^{-1}$ (left).
Lower panels: energy-scale dependence of {\tt MMHT14} NLO gluon and quark PDFs for $x= 10^{-2}$ (left) and $x = 10^{-4}$ (right).}
\label{fig:FFs_PDFs}
\end{figure}

\begin{figure}[t]
\centering

   \includegraphics[scale=0.50,clip]{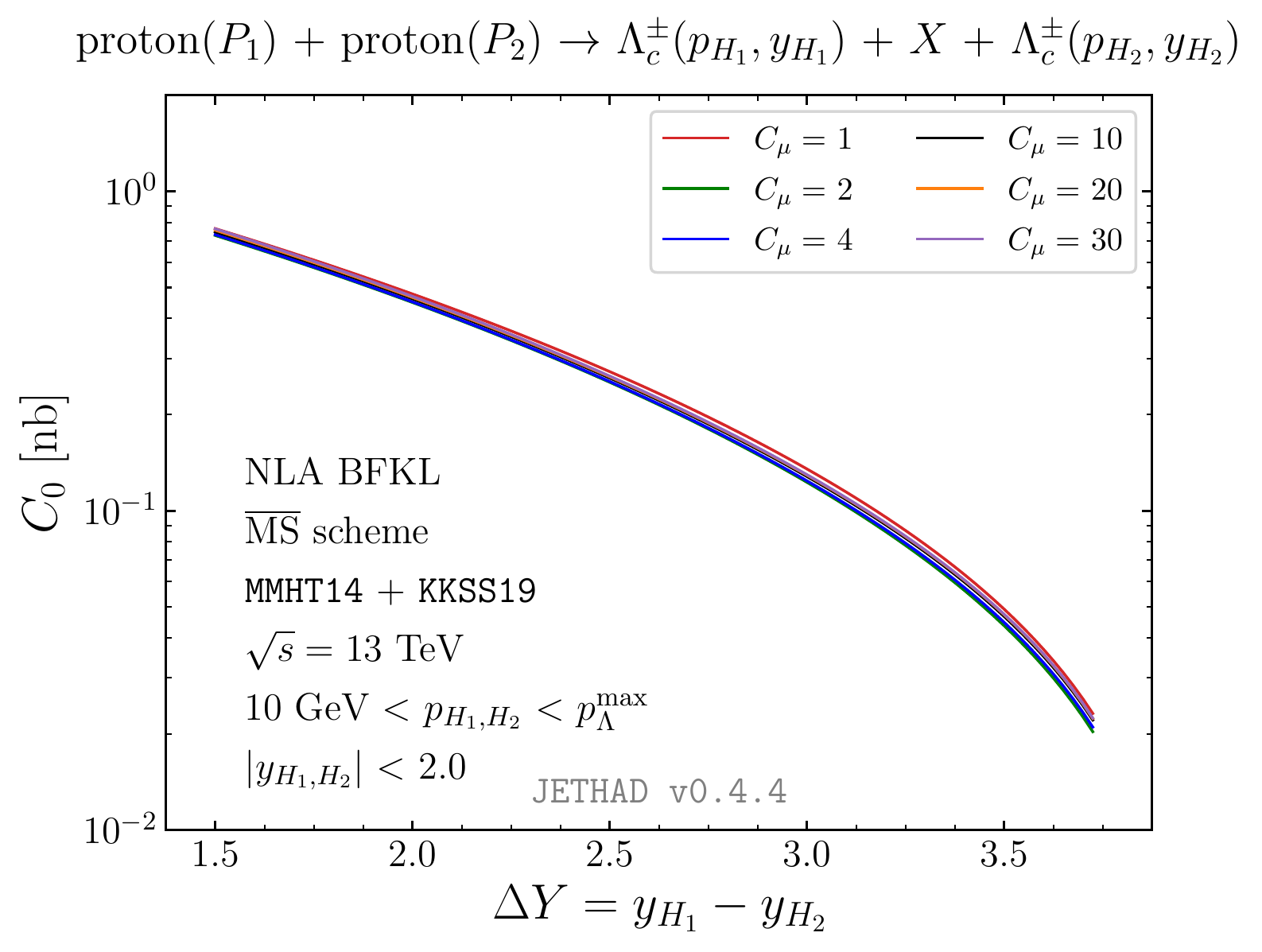}
   \includegraphics[scale=0.50,clip]{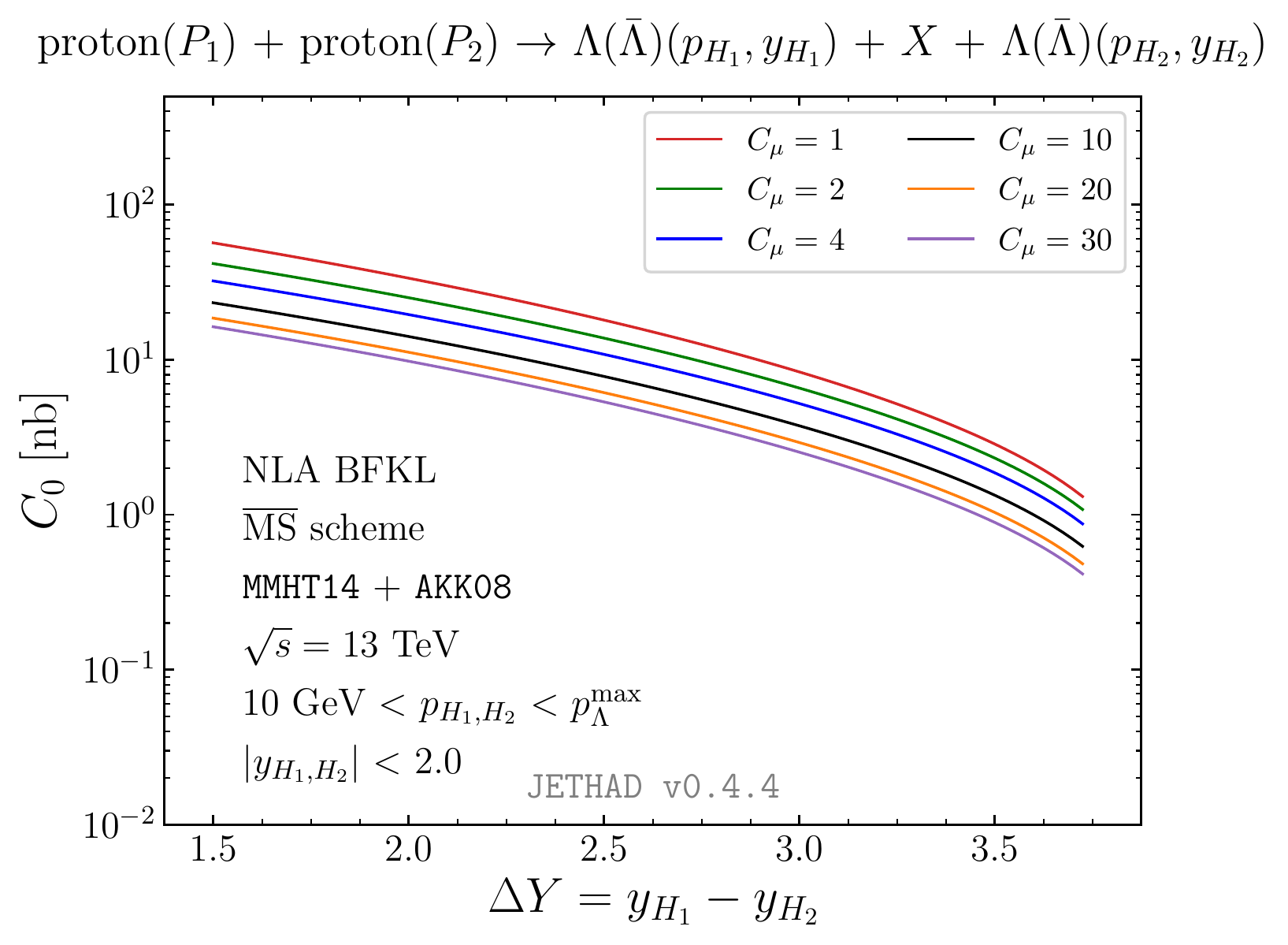}

   \includegraphics[scale=0.50,clip]{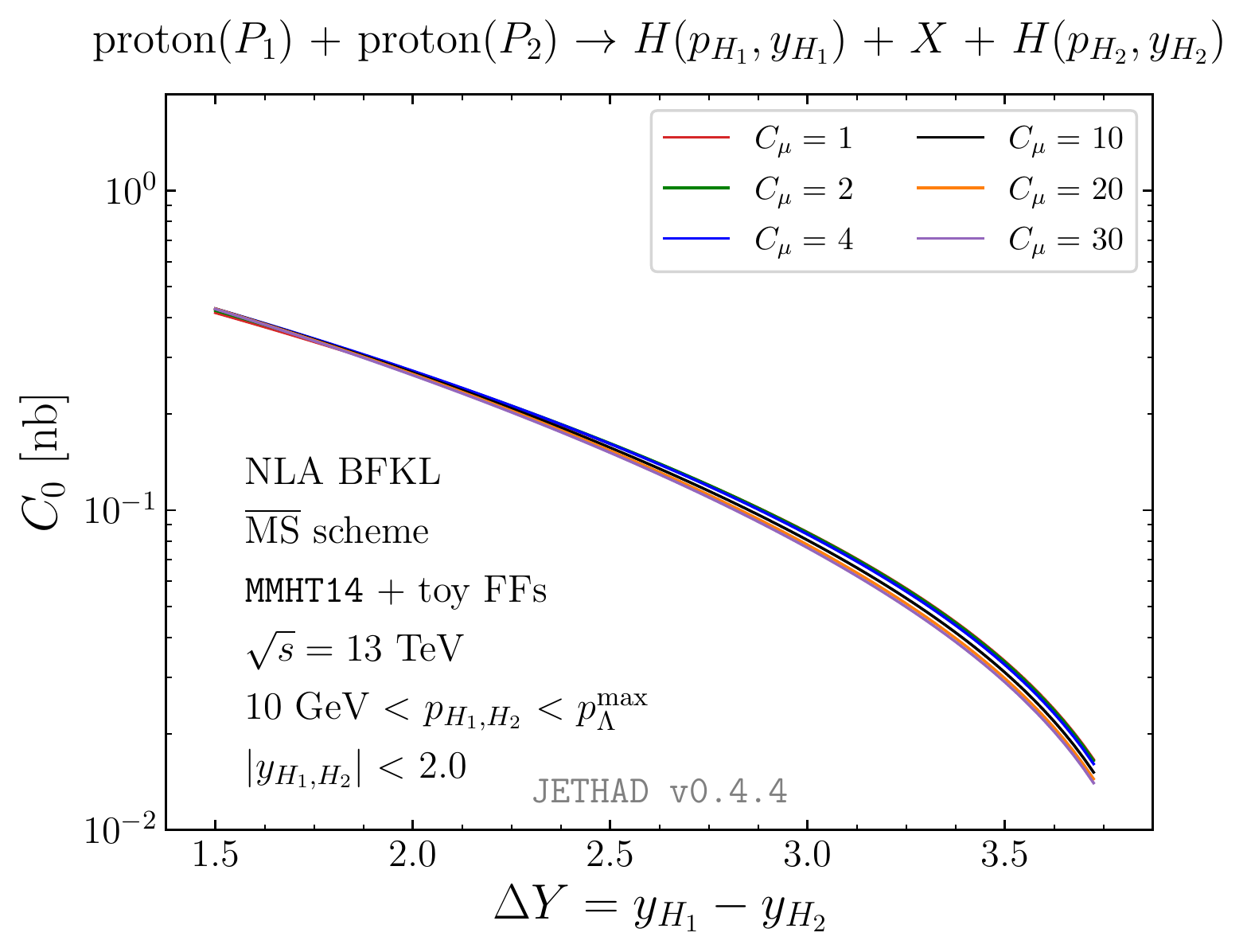}

\caption{Behavior of the $\varphi$-summed cross section, $C_0$, as a function of $\Delta Y$, in the dihadron production channel, and for $\sqrt{s} = 13$ TeV. 
A study on progressive energy-scale variation in the range $1 < C_{\mu} < 30$ is done for $\Lambda_c$ emissions (upper left), for $\Lambda$ detections (upper right), and for the production of a generic hadron species (lower) described by toy-FF parameterizations.}
\label{fig:C0_psv}
\end{figure}

In this Section we investigate the connection between $\Lambda_c$ FFs and the stability of cross sections under energy-scale variation. In upper panels of Fig.\tref{fig:FFs_PDFs} we compare the $\mu_F$-dependence of {\tt KKSS19} $\Lambda_c$ FFs (left) with {\tt AKK08} $\Lambda$ ones (right) for a value of the hadron momentum fraction typical of our analysis, namely $z = 0.5$. As expected, heavy flavors ($c$- and $b$-quarks) heavily dominate in $\Lambda_c$ fragmentation, whereas $s$-quark prevails in $\Lambda$ emission, here the lighter-quark species and the gluon giving however a more appreciable contribution. Notably, {\tt KKSS19} FFs smoothly increase with $\mu_F$ until they reach a constant value (apart from the $b$-quark, which decreases and then becomes constant). Conversely, {\tt AKK08} functions soften when $\mu_F$ raises. This difference turns out to be relevant when FFs are convoluted with PDFs in our LO impact factors~(Eq.~(\ref{LOLIF})). Since the dominant contribution to PDFs in the kinematic sector of our interest, where the longitudinal-momentum fraction $x$ ranges approximately from $10^{-4}$ to $10^{-2}$, is given by the gluon (see lower panels of Fig.\tref{fig:FFs_PDFs}), the behavior of the gluon FF also becomes relevant. Indeed, the employment of large scales, such as the BLM ones, gives rise to two competing effects. On the one side, larger $\mu_R$ values translate in a numerically smaller running coupling, both in the exponentiated kernel and the impact factors. On the other side, larger $\mu_F$ values heighten the gluon-PDF contribution. When this last is convoluted with a smooth-behaved, non-decreasing gluon FF, such as the $\Lambda_c$, the two features offset each other, thus generating the stability of cross sections under scale variation discussed in Section\tref{discussion}. \emph{Vice versa}, the downtrend with $\mu_F$ of the $\Lambda$-hyperon gluon FF spoils the balance between the two effects, thus preventing cross sections from reaching stability.
This statement is supported by predictions shown in Fig.\tref{fig:C0_psv}. Here, we test $C_0$ for double production of $\Lambda_c$ baryons (upper left panel) or $\Lambda$ hyperons (upper right panel) without applying the BLM method, under a progressive variation of energy-scales in a wider range that includes the typical BLM ones, $1 < C_\mu < 30$. We compare these results with corresponding ones for the inclusive emission, in the same kinematic domain, of two \emph{toy} hadrons (lower panel) described in term of the following, flavor and $\mu_F$-independent model of FFs
\begin{equation}
\label{toy_FFs}
 D_{g/q}^H (z, \mu_F)
 \; \to \;
 D^{\rm [toy]}(z) = {\cal N} z^{- \lambda} (1-z)^{3 \lambda} \; ,
\end{equation}
with ${\cal N} \simeq 1.5 \times 10^{-5}$ and $\lambda = 0.2$. The $z$-dependence of the model does not play a crucial role in our test. We clearly note that $C_0$ exhibits a fair stability under progressive scale variation both in the $\Lambda_c$ and in the toy-hadron channels, while its sensitivity spans over almost one order of magnitude in the hyperon case. This corroborates our assumption that smoothly-behaved, non-decreasing with $\mu_F$ FFs ($\mu_F$-constant, in the toy case) stabilize cross sections. Furthermore, the flavor independence of toy FFs removes any possible modulation on parton densities, thus confirming that the gluon channel drives the growth with $\mu_F$ of the convolution between PDFs and FFs.

The main objection against our statement could be that, at NLA, the large contribution of $c$- and $b$-quark FFs for $\Lambda_c$ is not anymore dampened by the smallness of intrinsic heavy-quark PDFs. Indeed, the term proportional to the $C_{gq}$ non-diagonal coefficient that appears in the hadron NLO impact-factor correction (see Eq.~(4.58) of Ref.\tcite{Ivanov:2012iv}) is enhanced by large heavy-flavor FFs, whose sum is in turn multiplied by the gluon PDF. Thus, the production channel that opens up at NLA can in principle compete with the pure LO one and its effect could spoil the description presented above. We numerically checked, however, that in kinematic ranges typical of our investigation the $C_{gg}$ diagonal coefficient strongly prevails over the non-diagonal ones, its regular part being larger 50 to $10^4$ times than $C_{gq}$. Therefore, gluon dynamics still dominates at NLA and this confirms our statement that the peculiar behavior of the $\Lambda_c$ gluon FF is responsible for the stability of our distributions under higher-order corrections.

\section{Summary and prospects}
\label{conclusions}

We studied the inclusive hadroproduction of a forward $\Lambda_c$ baryon in association with a backward object (another $\Lambda_c$ or a light jet) as a new probe channel of the semi-hard regime of QCD.
Results for rapidity-distance distributions and azimuthal-angle correlations, calculated within a hybrid factorization that combines collinear ingredients inside a full NLA BFKL treatment, offer a promising statistics and exhibit standard features of the high-energy dynamics.
We provided with a two-fold analysis on the sensitivity of energy scales, based on the variation of $\mu_R$ (and $\mu_F$) around their natural values as well as around the ones prescribed by the BLM optimization scheme.
We discovered that the tag of $\Lambda_c$ particles allows us to dampen the instabilities of the BFKL series, this resulting in a partial stabilization of resummed distributions under higher-order logarithmic corrections. This effect is more pronounced in the double $\Lambda_c$ production, while further studies on the $\Lambda_c$ plus jet channel will gauge the dependence of our observables on intrinsic features of the description of jet emissions, such as the selection algorithm. We plan to extend our program on semi-hard phenomenology by hunting for stabilizing effects via the inclusive production of heavier particles, such as $\Lambda_b$ baryons, heavy-light mesons and quarkonium states (see Refs.\tcite{Arbuzov:2020cqg,Chapon:2020heu} for a selection of phenomenological prospects in wider kinematic ranges, and Ref.\tcite{AbdulKhalek:2021gbh} for applications at the EIC). 

\section*{Acknowledgements}

We thank the Authors of Ref.\tcite{Kniehl:2020szu} for allowing us to link native {\tt KKSS19} FF routines to the {\tt JETHAD} code. We are indebted to Valerio Bertone for a discussion on heavy-flavor FFs.

F.G.C. acknowledges support from the INFN/NINPHA project and thanks the Universit\`a degli Studi di Pavia for the warm hospitality.
M.F. and A.P. acknowledge support from the INFN/QFT@COLLIDERS project.
The work of D.I. was carried out within the framework of the state contract of the Sobolev Institute of Mathematics (Project No. 0314-2019-0021).

\bibliographystyle{apsrev}
\bibliography{references}

\end{document}